\documentclass[journal]{IEEEtran}
\usepackage{amsmath,amsfonts}
\usepackage{algorithmic}
\usepackage{algorithm}
\usepackage{array}
\usepackage[caption=false,font=footnotesize,labelfont=rm,textfont=rm]{subfig}
\usepackage{textcomp}
\usepackage{stfloats}
\usepackage{url}
\usepackage{verbatim}
\usepackage{graphicx}
\usepackage{cite}
\usepackage{bm}
\usepackage{amssymb} 
\usepackage{mdwmath}
\usepackage{mdwtab}
\usepackage{eqparbox}
\usepackage{cuted}
\usepackage[hidelinks]{hyperref}
\usepackage{booktabs}

\usepackage{color}

\newtheorem{proposition}{Proposition}
\newtheorem{lemma}{Lemma}

\newtheorem{remark}{Remark}

\begin{document}

\title{Cell-Free MIMO with Rotatable Antennas: When Macro-Diversity Meets Antenna Directivity}

\author{Xingxiang Peng, Qingqing Wu, Ziyuan Zheng, Yanze Zhu, Wen Chen, Penghui Huang, \\Ying Gao, and Honghao Wang
\thanks{\textit{(Corresponding author: Qingqing Wu.)}}
\thanks{The authors are with the Department of Electronic Engineering, Shanghai Jiao Tong University, 200240, China
    (e-mail: 
    \href{mailto:peng_xingxiang@sjtu.edu.cn}{\nolinkurl{peng_xingxiang@sjtu.edu.cn}};
    \href{mailto:qingqingwu@sjtu.edu.cn}{\nolinkurl{qingqingwu@sjtu.edu.cn}};
    \href{mailto:zhengziyuan2024@sjtu.edu.cn}{\nolinkurl{zhengziyuan2024@sjtu.edu.cn}};
    \href{mailto:yanzezhu@sjtu.edu.cn}{\nolinkurl{yanzezhu@sjtu.edu.cn}};
    \href{mailto:wenchen@sjtu.edu.cn}{\nolinkurl{wenchen@sjtu.edu.cn}};
    \href{mailto:huangpenghui@sjtu.edu.cn}{\nolinkurl{huangpenghui@sjtu.edu.cn}};
    \href{mailto:yinggao@sjtu.edu.cn}{\nolinkurl{yinggao@sjtu.edu.cn}};
    \href{mailto:hhwang@sjtu.edu.cn}{\nolinkurl{hhwang@sjtu.edu.cn}}).
    }
}

\maketitle

\begin{abstract}
    Cell-free networks leverage distributed access points (APs) to achieve macro-diversity, yet their performance is often constrained by large disparities in channel quality arising from user geometry and blockages. To address this, rotatable antennas (RAs) add a lightweight hardware degree of freedom by steering the antenna boresight toward dominant propagation directions to strengthen unfavorable links, thereby enabling the network to better exploit macro-diversity for higher and more uniform performance. This paper investigates an RA-enabled cell-free downlink network and formulates a max-min rate problem that jointly optimizes transmit beamforming and antenna orientations. To tackle this challenging problem, we develop an alternating-optimization-based algorithm that iteratively updates the beamformers via a second-order cone program (SOCP) and optimizes the antenna orientations using successive convex approximation. To reduce complexity, we further propose an efficient two-stage scheme that first designs orientations by maximizing a proportional-fair log-utility using manifold-aware Frank-Wolfe updates, and then computes the beamformers using an SOCP-based design. Simulation results demonstrate that the proposed orientation-aware designs achieve a substantially higher worst-user rate than conventional beamforming-only benchmarks. Furthermore, larger antenna directivity enhances fairness with proper orientation but can degrade the worst-user performance otherwise.
\end{abstract}


\section{Introduction}
    \IEEEPARstart{T}{he} sixth-generation (6G) wireless network is expected to evolve toward greener, more flexible, and lightweight architectures while supporting ultra-high data rates, massive connectivity, and a uniformly high quality of experience \cite{intro1}. Achieving these targets requires reliable and spatially uniform service despite uneven link qualities, co-channel interference, and dynamic traffic loads. However, conventional cell-centric designs suffer from boundary effects and inter-cell disparities, making it challenging to guarantee consistent performance for cell-edge and disadvantaged users \cite{intro2,intro3,intro4}.

    To address these limitations, cell-free massive multiple-input multiple-output (MIMO) has emerged as a promising network architecture for beyond-5G and 6G systems \cite{intro5,intro6}. In a cell-free network, a large number of geographically distributed, low-complexity access points (APs) are connected to one or more central processing units (CPUs) via fronthaul links and cooperate to serve users in a non-cellular manner, thereby removing fixed cell boundaries. By allowing each user to benefit from multiple nearby APs, cell-free transmission provides macro-diversity and effectively mitigates cell-edge degradation, leading to more uniform service quality over the coverage area. Moreover, the reduced access distances and distributed cooperation can enhance spectral and energy efficiency, while offering greater flexibility in deployment and load balancing compared to conventional cellular architectures \cite{intro7,intro8,intro9}. Building on these architectural advantages, extensive research has investigated how to better harness cell-free cooperation through joint signal processing and resource optimization. In particular, coordinated beamforming and power control across distributed APs can markedly improve interference management, thereby reducing the transmit power required to meet a given quality-of-service target \cite{intro10,intro11}. Additional works have explored a range of performance objectives, including energy-efficiency optimization \cite{intro12}, spectral-efficiency maximization \cite{intro13}, and fairness-oriented formulations that aim to provide more consistent service quality across users \cite{intro14}.

    Beyond beamforming and power control, there is growing interest in further enhancing cell-free performance with limited additional hardware and energy overhead, toward greener and lightweight 6G deployments. This has spurred the incorporation of reconfigurable physical-layer technologies into the cell-free architecture, with representative examples including intelligent reflecting surfaces (IRSs) and movable antennas (MAs) \cite{R1,R2,R3}. An IRS is a nearly passive metasurface that applies programmable phase shifts to the incident wavefront, enabling low-cost control of the effective propagation environment \cite{intro15,intro16,R4}. In IRS-aided cell-free networks, the resulting reflected paths can strengthen weak or blocked AP--user links and facilitate more effective multi-AP cooperation, thereby complementing downlink beamforming in interference-limited regimes \cite{intro17,intro18,intro19}. Another representative technology is the MA architecture, where antenna elements are mounted on position-adjustable platforms and can be relocated within a prescribed region to adapt the array geometry to the surrounding propagation conditions \cite{intro20,intro21}. Unlike IRSs that reshape the incident fields through passive reflection, MAs directly modify the transmitter/receiver geometry and hence the effective channels by changing path lengths, phases, and spatial correlation, offering a more direct and flexible means to improve channel strength and orthogonality. In cell-free networks, such geometry adaptation can alleviate the imbalance of AP--user links and enhance channel separability across users, improving the efficiency of multi-AP cooperation and reducing the reliance on dense deployments \cite{intro22,intro23,intro24}.

    However, existing studies on MA-aided cell-free systems typically assume isotropic element patterns, whereas practical AP arrays often employ directional elements. How to exploit antenna directivity to further enhance cooperation and improve service uniformity therefore remains underexplored. The recently proposed six-dimensional movable antenna (6DMA) concept generalizes MAs by enabling joint adaptation of antenna positions and orientations \cite{intro25,intro26}, thereby introducing additional degrees of freedom for link alignment and interference mitigation \cite{intro27,intro271,intro272,intro28,intro29,intro30}. As an implementation-friendly subset of 6DMA, rotatable antennas (RAs) or rotatable surfaces optimize only the orientation dimension \cite{R5,R6,intro31,intro32,intro33,intro34,R8}. RAs can be realized either mechanically or electronically, both aiming to enable controllable steering of the main radiation direction of a directional element. Compared with position-reconfigurable MA and full 6DMA, RAs keep the element positions fixed, thus avoiding translation, aperture reconfiguration, collision management, and complex cable movement. This provides a simpler way to exploit orientation-domain reconfigurability with fewer mechanical degrees of freedom. RA-enabled designs have been investigated in applications such as integrated sensing and communication \cite{intro33}, physical-layer security \cite{intro34}, and spectrum sharing \cite{R8}. In particular, RAs are well suited to cell-free deployments with many distributed APs, where mechanical relocation and continuous position tracking can be costly, power-hungry, and slow. With directional elements, orientation control can reshape the effective channels and thus naturally complement network-wide beamforming in cell-free systems. Most recently, the work in \cite{intro35} incorporated RAs into cell-free networks, where each single-antenna AP is paired with a user and steers its boresight to improve the downlink performance. Yet, extending such an association-based design to a cell-free network with multiple RAs at each AP and fully cooperative multi-AP transmission is nontrivial and remains underexplored. In this case, the design variables change from AP--user association and one boresight vector per AP to network-wide beamformers together with element-wise RA orientations. More importantly, the RA orientations become tightly coupled with distributed beamforming through orientation-dependent channel gains, which fundamentally changes the problem structure. Consequently, neither the association-first decomposition nor the boresight-only optimization framework in \cite{intro35} is directly applicable. This motivates a new joint design to better exploit macro-diversity and unlock larger cooperation gains.

    Motivated by the above discussion, this paper investigates max--min fair downlink transmission in a cell-free network, where each AP is equipped with RAs whose boresights can be steered within a prescribed angular range. Compared with conventional cell-free systems with fixed antenna orientations, RA-enabled transmission can better exploit the available macro-diversity by adaptively reshaping the effective AP--user links. More importantly, the novelty here lies not in merely combining beamforming and RA optimization, but in revealing how antenna directivity reshapes AP--user link qualities and thereby changes how effectively macro-diversity can be exploited for fairness-oriented transmission. Specifically, RA steering harnesses antenna directivity to reinforce weak AP--user links while mitigating dominant interference, thereby reducing link-quality disparity and improving the cooperative support available to bottleneck users. As a result, RA can enhance distributed downlink beamforming beyond what is achievable with conventional beamforming alone, thus boosting the worst-user rate of the network. Overall, the main contributions of this work are summarized as follows:

    \begin{itemize}
        \item 
        We formulate a max--min rate maximization problem for RA-aided cell-free downlink transmission that jointly optimizes the beamformers and antenna orientations under per-AP power budgets and spherical-cap constraints. The resulting problem is highly nonconvex due to the coupling between beamforming and orientations, the orientation-dependent radiation gains embedded in the channels, and the unit-norm constraints.
    
        \item 
        We develop an alternating-optimization (AO) framework to tackle this nonconvex problem. With fixed RA orientations, the max--min beamforming subproblem is solved optimally via bisection, where each feasibility check reduces to a second-order cone program (SOCP). With fixed beamformers, the RA orientations are updated via successive convex approximation (SCA) with a relaxation of the unit-norm constraints, followed by a normalization step that restores feasibility without degrading the achieved minimum rate. Consequently, the AO iterations produce a non-decreasing minimum rate and are guaranteed to converge.
    
        \item 
        To reduce the computational complexity, we further propose a two-stage scheme. In Stage~1, the antenna orientations are designed by maximizing a proportional-fair log-utility of users' aggregate channel gains, which is optimized efficiently using manifold-aware Frank--Wolfe updates on the spherical-cap feasible set. In Stage~2, the beamformers are computed once using the SOCP-based max--min design with the obtained orientations.
    
        \item 
        Simulation results demonstrate that incorporating RA orientation optimization consistently improves the worst-user rate over beamforming-only baselines and random orientation schemes. A moderate steering range already captures most of the achievable gain, which supports practical actuator constraints. The results also reveal that antenna directivity is beneficial for fairness only when the boresights are properly optimized, while misaligned highly directional elements can degrade the worst-user performance. 
    \end{itemize}

    The remainder of this paper is organized as follows. Section~II presents the system model for RA-aided cell-free downlink transmission and formulates the max--min rate optimization problem. Section~III develops an AO-based algorithm to solve the formulated nonconvex problem. Section~IV proposes an efficient two-stage algorithm based on proportional-fair orientation design. Section~V provides numerical results and discussions, and Section~VI concludes the paper.
    
    \textit{Notations:} Bold lowercase, bold uppercase, and calligraphic letters denote vectors, matrices, and sets, respectively. $\mathbb{R}$ and $\mathbb{C}$ denote the sets of real and complex numbers. The operators $(\cdot)^{T}$, $(\cdot)^{H}$, and $(\cdot)^{*}$ denote transpose, Hermitian transpose, and complex conjugate, respectively. The Euclidean norm is denoted by $\|\cdot\|_{2}$, and the modulus of a complex scalar $x$ is denoted by $|x|$. The inner product of vectors $\bm{x}$ and $\bm{y}$ is written as $\langle \bm{x},\bm{y}\rangle=\bm{x}^{H}\bm{y}$. The real-part and imaginary-part operators are denoted by $\Re\{\cdot\}$ and $\Im\{\cdot\}$, respectively. $\bm I$ denotes the identity matrix of appropriate size. The notation $\bm A\succeq \bm 0$ means that $\bm A$ is positive semidefinite, and $\bm A\preceq \bm B$ means that $\bm B-\bm A$ is positive semidefinite. The operator $[x]_{+}\triangleq \max\{0,x\}$. The natural and base-2 logarithms are denoted by $\ln(\cdot)$ and $\log_{2}(\cdot)$, respectively, while $e$ and $j\triangleq \sqrt{-1}$ denote Euler’s number and the imaginary unit. The distribution $\mathcal{CN}(0,\sigma^{2})$ denotes a circularly symmetric complex Gaussian random variable with zero mean and variance $\sigma^{2}$, and $\mathcal{U}[a,b]$ denotes the continuous uniform distribution over $[a,b]$. The operator $\operatorname{det}(\cdot)$ denotes the determinant, and $\mathrm{SO}(3)$ denotes the special orthogonal group in three dimensions.

\begin{figure}[t]
	\begin{center}
		\includegraphics[width=0.48\textwidth]{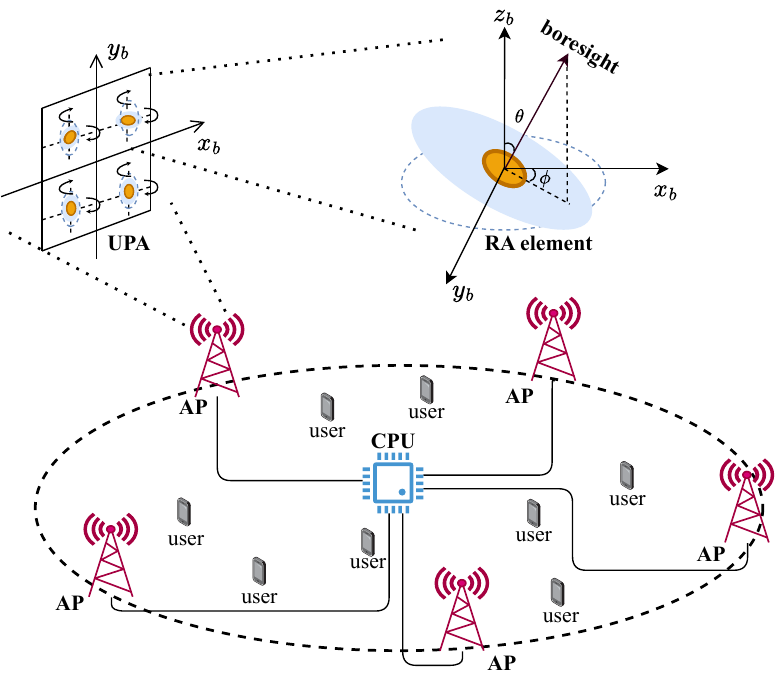}
		\caption{A cell-free MIMO network with rotatable antennas.}
		\label{fig1}
	\end{center}
\end{figure}

\section{System Model and Problem Formulation}
As shown in Fig.~\ref{fig1}, we consider an RA-aided cell-free downlink network comprising $B$ access points (APs) and $K$ users. Each user is equipped with a fixed isotropic antenna\footnote{This assumption is motivated by practical user devices, which are typically size-, cost-, and power-constrained and are therefore commonly modeled as fixed isotropic antennas.}, while each AP $b\in\{1,\ldots,B\}$ employs a uniform planar array (UPA) of $M$ rotatable antennas (RAs). The UPA lies in the AP's local $x_b$--$y_b$ plane, where the $M$ RAs are arranged in an $M_x\times M_y$ grid with uniform inter-element spacing $d$ along both axes (i.e., $M=M_xM_y$). Using row-major indexing, the RA at local grid coordinate $(m_x,m_y)$, with $m_x\in\{1,\ldots,M_x\}$ and $m_y\in\{1,\ldots,M_y\}$, is assigned the index $m\in\{1,\ldots,M\}$ as 
\begin{align}
    m &= m_{x} + (m_{y}-1)M_{x}. \label{eq:rowmajor1}
\end{align}
The local position of antenna element $m$, relative to the UPA center, is given by
\begin{align}
  \tilde{\bm{p}}_{b,m}
  &= \begin{bmatrix}
       \left(m_{x}-\frac{M_{x}+1}{2}\right)d \\
       \left(m_{y}-\frac{M_{y}+1}{2}\right)d \\
       0
     \end{bmatrix},
\end{align}
where the origin is at the panel's geometric center and the array normal aligns with the local $z_b$-axis. Let $\bm{R}_b\in\mathrm{SO}(3)$ denote the rotation matrix from the local coordinate system (LCS) $(x_b,y_b,z_b)$ to the global coordinate system (GCS), and let $\bm{p}_{b,0}\in\mathbb{R}^3$ denote the global coordinate of the UPA center. Then, the global position of antenna element $m$ is given by
\begin{align}
  \bm{p}_{b,m}
  &= \bm{p}_{b,0} + \bm{R}_b\,\tilde{\bm{p}}_{b,m}.
  \label{eq:global_pos}
\end{align}
Here, $\bm R_b$ is constructed by stacking the three local basis vectors of AP $b$ expressed in the GCS, i.e.,
\begin{align}
    \bm R_b = [\bm e_{x_b},\bm e_{y_b},\bm e_{z_b}],
\end{align}
where $\bm e_{x_b}$, $\bm e_{y_b}$, and $\bm e_{z_b}$ are the three right-handed orthonormal basis vectors of the LCS of AP $b$, expressed in the GCS.

\subsection{RA Model}
We assume that the position and pose of each UPA remain fixed, i.e., the rotation matrix $\{\bm{R}_b\}_{b=1}^{B}$ is predetermined during deployment. However, each antenna element on the UPA can independently adjust its orientation through electronic or mechanical means. For UPA $b$, the orientation of element $m$ is parameterized in its LCS by the zenith angle $\theta_{b,m}$, measured from the local $z_b$-axis, and the azimuth angle $\phi_{b,m}$, measured from the local $x_b$-axis in the $x_b$–$y_b$ plane, expressed as
\begin{align}
  \tilde{\bm{f}}_{b,m}(\theta_{b,m},\phi_{b,m})
  &= \begin{bmatrix}
       \sin\theta_{b,m}\cos\phi_{b,m} \\
       \sin\theta_{b,m}\sin\phi_{b,m} \\
       \cos\theta_{b,m}
     \end{bmatrix},
\end{align}
which inherently satisfies $\|\tilde{\bm{f}}_{b,m}\|_{2} = 1$ by construction. To account for mechanical constraints and mitigate mutual coupling between adjacent elements, the zenith angle is constrained by
\begin{align}
    \label{fconstr}
    0 \le \theta_{b,m} \le \theta_{\max}, \quad \forall b,m,
\end{align}
where $\theta_{\max} \in [0,\tfrac{\pi}{2})$. This angular constraint on $\theta_{b,m}$ can be equivalently expressed as a constraint on the boresight vector:
\begin{align}
    \cos(\theta_{\max}) \le \tilde{\bm{f}}_{b,m}^{T}\bm{e}_z \le 1, \quad \forall b,m,
\end{align}
where $\bm{e}_z = [0,0,1]^T$ represents the unit vector along the $z$-axis in the LCS. This equivalence follows from $\tilde{\bm f}_{b,m}^{T}\bm e_z=\cos\theta_{b,m}$, and this vector form is convenient for optimizing the RA orientations over the spherical-cap feasible set. The corresponding boresight vector in the GCS is then given by
\begin{align}
  \bm{f}_{b,m} = \bm{R}_b\,\tilde{\bm{f}}_{b,m}.
\end{align}
Here, $\bm R_b$ serves only as a fixed LCS-to-GCS coordinate transformation, and therefore does not change the local boresight definition or the subsequent orientation-dependent channel expressions. We model each RA element as a directional radiator. The directional gain of each RA depends on the angular offset $\epsilon$ between the signal direction and the element's main-lobe boresight, and can be modeled by a cosine pattern \cite{nr1,R8}:
\begin{align}
    G(\epsilon) =
    \begin{cases}
        \kappa_{\text{max}} \cos^{2p}(\epsilon), & \epsilon \in \left[0, \tfrac{\pi}{2}\right], \\[4pt]
        0, & \text{otherwise},
    \end{cases}
    \label{eq:RA_pattern}
\end{align}
where $\epsilon$ denotes the angular offset between the signal direction and the antenna boresight, and $p$ is the antenna directivity parameter. The coefficient $\kappa_{\max}=2(2p+1)$ is used for power normalization so that the total radiated power remains consistent when changing $p$. Within this cosine-power model, a larger $p$ corresponds to a narrower and more concentrated main lobe, and hence $p$ serves as a key parameter for characterizing antenna directivity.

\subsection{Channel Model}
We adopt a narrowband far-field model for the AP–user channels. Let $\bm{u}_k\in\mathbb{R}^3$ denote the global position of user $k$. Using the cosine-pattern gain $G(\epsilon)$ in \eqref{eq:RA_pattern} and Friis’ law, the line-of-sight (LoS) link-power gain from element $(b,m)$ to user $k$ can be expressed as \cite{nr1}
\begin{align}
  G^{\mathrm{LoS}}_{k,b,m}
  &= \beta_0\,r_{k,b,m}^{-2}\,G(\epsilon_{k,b,m}) \notag
   \\
   &=\; \beta_0\,r_{k,b,m}^{-2}\,\kappa_{\text{max}}\,[( \bm{R}_b\,\tilde{\bm{f}}_{b,m})^T\bm{s}_{k,b,m}]_+^{2p},
  \label{eq:los_power}
\end{align}
where $\lambda$ denotes the carrier wavelength, $\beta_0=(\lambda/4\pi)^2$ is the free-space path-power gain at the reference distance of 1 meter, $r_{k,b,m} = \|\bm{p}_{b,m}-\bm{u}_k\|_2$, and $\bm{s}_{k,b,m} = \frac{\bm{u}_k-\bm{p}_{b,m}}{r_{k,b,m}}$ is the unit vector pointing from element $(b,m)$ to user $k$. The corresponding LoS channel coefficient is given by
\begin{align}
  h^{\mathrm{LoS}}_{k,b,m}
  &= \sqrt{G^{\mathrm{LoS}}_{k,b,m}}\;
     e^{-j\frac{2\pi}{\lambda}r_{k,b,m}}.
  \label{eq:los_coeff}
\end{align}
Consider $Q$ scatterers at $\{\bm{c}_q\}_{q=1}^Q$ with radar cross-section (RCS) $\{\zeta_q\}_{q=1}^Q$. Define the element–scatterer distances and directions as $\tilde{r}_{q,b,m} = \|\bm{c}_q-\bm{p}_{b,m}\|_2$, and $\bm{s}_{q,b,m} = \frac{\bm{c}_q-\bm{p}_{b,m}}{\tilde{r}_{q,b,m}}$. Then the link-power gain from element $(b,m)$ to scatterer $q$ is given by
\begin{align}
  G_{q,b,m}
  &= \beta_0\,\tilde{r}_{q,b,m}^{-2}\,\kappa_{\max}\,[( \bm{R}_b\,\tilde{\bm{f}}_{b,m})^T\bm{s}_{q,b,m}]_+^{2p}.
\end{align}
Under a bistatic scattering model, the non-line-of-sight (NLoS) channel coefficient from element $(b,m)$ to user $k$ is given by \cite{nr1}
\begin{align}
  h^{\mathrm{NLoS}}_{k,b,m}
  &= \sum_{q=1}^{Q}
     \sqrt{\frac{\zeta_q\,G_{q,b,m}}{4\pi\,\hat r_{k,q}^2}}\;
     e^{-j\frac{2\pi}{\lambda}(\tilde r_{q,b,m}+\hat r_{k,q})+j\chi_q},
  \label{eq:nlos_coeff}
\end{align}
where $\hat r_{k,q}  = \|\bm{u}_k-\bm{c}_q\|_2$ is the scatterer-user distance, and $\chi_q\sim\mathcal{U}[0,2\pi)$ denotes an independent random phase. Accordingly, the total coefficient from element $(b,m)$ to user $k$ is given by
\begin{align}
    h_{k,b,m} &= h^{\mathrm{LoS}}_{k,b,m} + h^{\mathrm{NLoS}}_{k,b,m},
\end{align}
and the per-AP channel vector is expressed as
\begin{align} \label{chanCom}
  \bm{h}_{k,b}
  &= \big[h_{k,b,1},\cdots,h_{k,b,M}\big]^T \in \mathbb{C}^{M}.
\end{align}

\begin{remark}[Element-wise orientation vs. UPA pose control]
    In this work, we focus on element-wise RA orientation control while keeping the UPA pose matrices $\{\bm{R}_b\}_{b=1}^B$ fixed across the network deployment. Under this assumption, the antenna orientations affect the channel only through the directional gain terms $G(\epsilon)$ in \eqref{eq:RA_pattern}, while the path lengths and associated phase shifts are governed by the static array and user/scatterer locations. This modeling provides finer-grained flexibility than UPA-wise rotation, since different elements can adjust their boresights to serve users in different spatial directions, but it may also incur higher implementation and calibration complexity. On the other hand, allowing the UPA poses $\{\bm{R}_b\}_{b=1}^B$ themselves to be reconfigurable would simultaneously change both the element positions and the array boresights, leading to joint variations in path-lengths, phases, and power patterns. Such a fully pose-adaptive architecture entails a more involved system model and is left for future work.
\end{remark}

\subsection{Signal Model}

In the considered cell-free architecture, all APs are connected via high-capacity fronthaul links to a central processing unit (CPU). The CPU performs joint baseband processing and downlink precoding based on global channel state information (CSI). Hence, each user's data symbol can be made available at all APs to enable coherent joint transmission. The transmit signal from AP $b$ is given by
\begin{align}
  \bm{x}_b
  &= \sum_{k=1}^{K}\bm{w}_{k,b}\,s_k,
  \label{eq:txsig}
\end{align}
where $s_k\sim\mathcal{CN}(0,1)$ denotes the data symbol intended for user $k$, and $\bm{w}_{k,b}\in\mathbb{C}^{M}$ represents the corresponding beamforming vector at AP $b$. The received signal at user $k$ can then be expressed as
\begin{align}
  y_k
  &= \sum_{b=1}^{B}\bm{h}_{k,b}^H\,\bm{w}_{k,b} s_k
   + \sum_{b=1}^{B}\sum_{ j\neq k}^{K}\bm{h}_{k,b}^H\,\bm{w}_{j,b}\,s_j
   + n_k,
  \label{eq:rxsig}
\end{align}
where $n_k\sim\mathcal{CN}(0,\sigma_k^2)$ denotes the additive white Gaussian noise (AWGN) at user $k$. For notational compactness, we stack the beamforming vectors across all APs for user $k$ as $\bm{w}_k= \big[\bm{w}_{k,1}^T,\cdots,\bm{w}_{k,B}^T\big]^T$, and similarly stack the corresponding channels as $\bm{h}_k = \big[\bm{h}_{k,1}^T,\cdots,\bm{h}_{k,B}^T\big]^T$. Then the received signal at user $k$ can be reformulated as
\begin{align}
  y_k
  &=\bm{h}_k^H\bm{w}_k\,s_k
       + \sum_{j\neq k}\bm{h}_k^H\bm{w}_j\,s_j + n_k.
  \label{eq:rxsig_compact}
\end{align}
Consequently, the achievable data rate for user $k$ is given by
\begin{align}
  R_k
  &=\; \log_2\!\left(1+ \frac{\bm{h}_k^H\bm{w}_k\bm{w}_k^H\bm{h}_k}{\bm{h}_k^H\!\!\sum_{j\neq k}\bm{w}_j\bm{w}_j^H\bm{h}_k+\sigma_k^2}\right),
  \label{eq:rate}
\end{align}
where the term inside the logarithm is the signal-to-interference-plus-noise ratio (SINR) at user $k$.

\subsection{Problem Formulation}
We aim to jointly design the RA element orientations and downlink beamformers to maximize the minimum user rate, thereby promoting fairness among users. Since $R_k$ in \eqref{eq:rate} is a strictly increasing function of the received SINR, this max–min rate problem is equivalently recast as a max–min SINR problem. Introducing an auxiliary variable $\gamma$ to represent the worst-case user SINR, the resulting optimization problem can be written as
\begin{align}
  \max_{\{\tilde{\bm{f}}_{b,m}\},\{\bm{w}_{k,b}\},\,\gamma} \,\,
  & \log_2(1+\gamma) \label{prob:mm_obj}\\
  \text{s.t.}\qquad\,\,
  &  \frac{\bm{h}_k^H\bm{w}_k\bm{w}_k^H\bm{h}_k}
  {\bm{h}_k^H\Big(\sum_{j\neq k}\bm{w}_j\bm{w}_j^H\Big)\bm{h}_k+\sigma_k^2} \ge \gamma,\,\, \forall k, \label{prob:mm_rate}\\
  & \sum_{k=1}^{K}\|\bm{w}_{k,b}\|_2^2 \,\le\, P_{\mathrm{max}}, \,\, \forall b,\label{prob:sum_power}\\
  & \cos(\theta_{\max}) \le \tilde{\bm{f}}_{b,m}^{T}\bm{e}_z \le 1, \,\, \forall b,m, \label{prob:orient_cap} \\
  & \|\tilde{\bm{f}}_{b,m} \|_2^2 = 1, \,\,  \forall b,m.  \label{prob:orient_norm}
\end{align}
Here, \eqref{prob:sum_power} imposes a per-AP transmit power budget $P_{\mathrm{max}}$, while \eqref{prob:orient_cap} and \eqref{prob:orient_norm} ensure that each RA element’s boresight lies on a spherical cap around the local $z$-axis with maximum zenith angle $\theta_{\max}$. The above max–min design problem is difficult to solve optimally for two main reasons. First, the beamformers and orientation vectors are intricately coupled through the SINR constraints \eqref{prob:mm_rate}, leading to a highly non-convex optimization problem. Second, the spherical-cap orientation constraints \eqref{prob:orient_cap}–\eqref{prob:orient_norm} render the feasible set non-convex, which further complicates the design. From a system-level perspective, antenna orientation reshapes the effective AP--user channel gains through the directional-gain terms. This provides an additional degree of freedom to strengthen weak AP--user links and improve the cooperative support for bottleneck users, thereby helping the cell-free network better exploit macro-diversity for fairness-oriented transmission.

\begin{remark}
The present formulation adopts an element-wise RA model together with an idealized orientation-aware antenna/channel model as a fine-grained benchmark for studying the role of RA orientation in cooperative cell-free transmission. In practical deployments, however, fully independent rotation may incur considerable overhead when the number of APs or antenna elements is large, and the adopted model does not explicitly capture higher-order effects such as pattern distortion, mutual coupling, spatial correlation, near-field propagation, polarization mismatch, or channel uncertainty. Accordingly, the reported results should be regarded as benchmark-type performance evaluations under idealized assumptions. More practical orientation-control architectures, as well as more realistic hardware-aware and robust RA designs, are important directions for future research.
\end{remark}

For ease of readability, the main system parameters used in the subsequent algorithm designs are summarized in Table~\ref{tab:main_parameters}.

\begin{table}[ht]
\renewcommand{\arraystretch}{0.9}
\caption{List of Main System Parameters}
\label{tab:main_parameters}
\centering
\begin{tabular}{c||l}
\toprule[1.5pt]
\textbf{Symbols} & \textbf{Descriptions} \\
\hline
$B, K$ & Numbers of APs and users \\
$M, M_x, M_y$ & Number of RAs per AP and UPA dimensions \\
$d$ & Inter-element spacing \\
$\bm p_{b,0}, \bm p_{b,m}$ & UPA-center and element positions of AP $b$ in the GCS \\
$\tilde{\bm p}_{b,m}$ & Element position in the LCS of AP $b$ \\
$\bm R_b$ & LCS-to-GCS rotation matrix of AP $b$ \\
$\bm u_k$ & Position of user $k$ in the GCS \\
$\tilde{\bm f}_{b,m}, \bm f_{b,m}$ & Local and global boresight vectors of element $(b,m)$ \\
$\theta_{b,m}, \phi_{b,m}$ & Zenith and azimuth angles of $\tilde{\bm f}_{b,m}$ \\
$\theta_{\max}$ & Maximum zenith angle of each RA \\
$p, \kappa_{\max}$ & Directivity parameter and peak gain coefficient \\
$\lambda, \beta_0$ & Carrier wavelength and reference path gain \\
$Q, \bm c_q$ & Number and position of scatterers \\
$\zeta_q, \chi_q$ & RCS and random phase of scatterer $q$ \\
$\bm h_{k,b}, \bm h_k$ & Per-AP and stacked channel vectors for user $k$ \\
$\bm w_{k,b}, \bm w_k$ & Per-AP and stacked beamforming vectors for user $k$ \\
$\sigma_k^2$ & Noise power at user $k$ \\
$P_{\max}$ & Transmit power budget of each AP \\
\bottomrule[1.5pt]
\end{tabular}
\end{table}

\section{Proposed AO-Based Solution}
This section develops an AO-based algorithm to solve the coupled max--min SINR problem. With fixed RA orientations, the beamformer subproblem is solved globally via bisection over the target SINR, where each feasibility check reduces to an SOCP. With fixed beamformers, we address the nonconvex orientation subproblem via an SCA method with a convex relaxation and a normalization step. Finally, we establish monotone convergence of the resulting AO iterations and analyze the computational complexity.

\subsection{Optimization of Beamformers}
\label{sec:beamforming}
Recall that $\bm w_k = [\bm w_{k,1}^T,\ldots,\bm w_{k,B}^T]^T$ denotes the stacked beamformer for user $k$. With the RA orientations fixed, the channels $\{\bm h_k\}_{k=1}^K$ are determined, and the max–min fair beamforming problem reduces to
\begin{align}
  \max_{\{\bm{w}_{k,b}\},\,\gamma}\quad  &\log_2(1+\gamma) \label{SOCPBEAM}\\
  \text{s.t.}\qquad
  &\frac{|\bm h_k^H\bm w_k|^2}{\sum_{j\neq k}|\bm h_k^H\bm w_j|^2+\sigma_k^2}\ \ge\gamma,\ \forall k, \label{eq:SINR}\\
  & \sum_{k=1}^K \|\bm w_{k,b}\|_2^2 \le P_{\max},\ \forall b. \label{eq:perBS}
\end{align}
Exploiting the phase invariance of the SINR, we can, without loss of optimality, apply a phase rotation to $\bm w_k$ such that $\Im\{\bm h_k^H\bm w_k\}=0$ and $\Re\{\bm h_k^H\bm w_k\}\ge 0$. Therefore, for any target SINR $\bar\gamma>0$, the constraints \eqref{eq:SINR} can be equivalently written as the following second-order cone (SOC) constraints:
\begin{align}
  \left\|
  \begin{bmatrix}
    \bm h_k^H\bm w_1\\[-1pt]
    \vdots\\[-1pt]
    \bm h_k^H\bm w_{k-1}\\[-1pt]
    \bm h_k^H\bm w_{k+1}\\[-1pt]
    \vdots\\[-1pt]
    \bm h_k^H\bm w_K\\
    \sigma_k
  \end{bmatrix}
  \right\|_2
  \ \le\ \frac{1}{\sqrt{\bar\gamma}}\ \Re\{\bm h_k^H\bm w_k\}.
  \label{eq:SOC}
\end{align}
This transformation shows that, for any fixed target SINR $\bar\gamma$, the beamformer design reduces to a convex SOCP feasibility problem, which can be efficiently solved using standard tools such as CVX \cite{nr2}. Consequently, the globally optimal solution to the beamforming problem \eqref{SOCPBEAM} can be obtained via a bisection search over $\bar\gamma$, where each iteration solves the corresponding SOCP to check feasibility.

\subsection{Optimization of RA Orientations}
With the beamformers $\{\bm w_k\}$ fixed, the RA orientation optimization problem is given by
\begin{align}
    \max_{\{\tilde{\bm{f}}_{b,m}\},\gamma} \quad & \log_2(1+\gamma) \label{prob:antenna_obj1} \\
    \text{s.t.} \qquad 
    & \frac{\bm{h}_k^H \bm{Q}_k \bm{h}_k}{\bm{h}_k^H \bm{Q}_{-k} \bm{h}_k + \sigma_k^2} \geq \gamma, \quad \forall k, \label{prob:antenna_rate} \\
    & \cos(\theta_{\max}) \leq \tilde{\bm{f}}_{b,m}^T \bm{e}_z \leq 1, \quad \forall b, m, \label{prob:antenna_cap} \\
    & \|\tilde{\bm{f}}_{b,m}\|_2^2 = 1, \quad \forall b, m, \label{prob:antenna_norm}
\end{align}
where $\bm Q_k \triangleq \bm w_k \bm w_k^H$ and $\bm Q_{-k} \triangleq \sum_{j\ne k} \bm w_j \bm w_j^H$ denote the signal and interference covariance matrices, respectively. Note that, under the adopted propagation model, the effective channels $\{\bm h_k\}$ depend on the RA orientations $\{\tilde{\bm f}_{b,m}\}$. To decouple the fractional SINR constraints in \eqref{prob:antenna_rate}, we introduce auxiliary variables $\{z_k>0\}$ and obtain the following equivalent reformulation:
\begin{align}
    \max_{\{\tilde{\bm{f}}_{b,m}\},\gamma,\{z_k\}} \quad & \log_2(1+\gamma) \label{prob:antenna_obj12e1_joint} \\
    \text{s.t.} \qquad \,\,
    & \bm{h}_k^H \bm{Q}_k \bm{h}_k  \geq z_k\gamma, \quad \forall k, \label{prob:antenna_rate34_joint} \\
    & \bm{h}_k^H \bm{Q}_{-k} \bm{h}_k + \sigma_k^2 \leq z_k,\,\, \forall k, \label{prob:antenna_rate34_joidnt}\\
    & \cos(\theta_{\max}) \leq \tilde{\bm f}_{b,m}^T \bm{e}_z \leq 1, \quad \forall b,m, \label{prob:antenna_c23ap_joint} \\
    & \|\tilde{\bm f}_{b,m}\|_2^2 = 1,  \quad \forall b,m.  \label{prob:antenna_4enordfm_joint}
\end{align}
The equivalence between \eqref{prob:antenna_rate} and \eqref{prob:antenna_rate34_joint}--\eqref{prob:antenna_rate34_joidnt} can be readily verified by contradiction. Nevertheless, the resulting problem remains highly nonconvex due to (i) the bilinear terms $z_k\gamma$ in \eqref{prob:antenna_rate34_joint}, (ii) the nonconvex dependence of the signal and interference powers $\bm h_k^H \bm Q_k \bm h_k$ and $\bm h_k^H \bm Q_{-k} \bm h_k$ on the orientation variables ${\tilde{\bm f}_{b,m}}$, and (iii) the unit-norm constraint in \eqref{prob:antenna_4enordfm_joint}, which renders the feasible set nonconvex. To address these issues, we adopt an SCA framework that solves a sequence of convex surrogate problems. As a first step, we upper-bound the bilinear terms $z_k\gamma$ by convex quadratic functions using Young’s inequality:
\begin{align}
    z_k\gamma \leq \frac{1}{2}\left(\frac{z_k^{[t]}}{\gamma^{[t]}}\gamma^2+\frac{\gamma^{[t]}}{z_k^{[t]}}z_k^2\right),\quad \forall k,
\end{align}
where $\gamma^{[t]}$ and $\{z_k^{[t]}\}$ are the values obtained at the $t$-th SCA iteration. Substituting these upper bounds for $z_k\gamma$ in \eqref{prob:antenna_rate34_joint} yields the following intermediate surrogate problem:
\begin{align}
    \max_{\{\tilde{\bm{f}}_{b,m}\},\gamma,\{z_k\}} \quad & \log_2(1+\gamma)  \\
    \text{s.t.} \qquad 
    & \bm{h}_k^H \bm{Q}_k \bm{h}_k \geq \frac{1}{2}\Big(\frac{z_k^{[t]}}{\gamma^{[t]}}\gamma^2+\frac{\gamma^{[t]}}{z_k^{[t]}}z_k^2\Big), \,\, \forall k, \label{dds1} \\
    & \bm{h}_k^H \bm{Q}_{-k} \bm{h}_k + \sigma_k^2 \leq z_k,\,\, \forall k, \label{dds2}\\
    & \cos(\theta_{\max}) \leq \tilde{\bm f}_{b,m}^T \bm{e}_z \leq 1,  \quad \forall b,m, \\
    & \|\tilde{\bm f}_{b,m}\|_2^2 = 1, \quad \forall b,m. \label{normCons}
\end{align}
Next, we construct quadratic surrogates for the signal and interference power functions in constraints \eqref{dds1} and \eqref{dds2}
\begin{align}\label{SigPNoisP}
\left\{
\begin{aligned}
    S_k(\{\tilde{\bm{f}}_{b,m}\}) &\triangleq \bm{h}_k^H \bm{Q}_k \bm{h}_k,\\
    I_k(\{\tilde{\bm{f}}_{b,m}\}) &\triangleq \bm{h}_k^H \bm{Q}_{-k} \bm{h}_k.
\end{aligned}
\right.
\end{align}
The functions $S_k(\{\tilde{\bm{f}}_{b,m}\})$ and $I_k(\{\tilde{\bm{f}}_{b,m}\})$ are neither convex nor concave in $\{\tilde{\bm{f}}_{b,m}\}$, so a first-order Taylor expansion alone cannot yield global lower or upper bounds. To this end, we resort to the second-order Taylor expansion with suitably chosen curvature parameters.

\begin{proposition}[Surrogate functions for $S_k$ and $I_k$] \label{prop1}
Suppose that the antenna directivity factor satisfies $p \ge 2$. Then, for each user $k$, there exists a finite constant $\xi_k>0$ such that $S_k$ admits the following global quadratic lower bound:
\begin{align}
    S_k(\{\tilde{\bm f}_{b,m}\})
    &\geq S_k(\{\tilde{\bm f}_{b,m}^{[t]}\}) \notag \\
    &\quad + \sum_{b,m}\nabla_{\tilde{\bm f}_{b,m}} S_k(\{\tilde{\bm f}_{b,m}^{[t]}\})^T\big( \tilde{\bm f}_{b,m}-\tilde{\bm f}_{b,m}^{[t]} \big)\notag \\
    &\quad -\frac{\xi_{k}}{2}\sum_{b,m}\big\| \tilde{\bm f}_{b,m}-\tilde{\bm f}_{b,m}^{[t]} \big\|_2^2 \notag \\
    &\triangleq \underline{S_k}^{[t]}(\{\tilde{\bm f}_{b,m}\}), \label{eqq11}
\end{align}
where $\{\tilde{\bm f}_{b,m}^{[t]}\}$ denotes the current SCA iterate and is treated as fixed when constructing the surrogate function. Similarly, there exists a finite constant $\chi_k>0$ such that $I_k$ admits the following global quadratic upper bound:
\begin{align}
    I_k(\{\tilde{\bm f}_{b,m}\}) 
    &\leq I_k(\{\tilde{\bm f}_{b,m}^{[t]}\}) \notag \\
    &\quad + \sum_{b,m}\nabla_{\tilde{\bm f}_{b,m}} I_k(\{\tilde{\bm f}_{b,m}^{[t]}\})^T\big( \tilde{\bm f}_{b,m}-\tilde{\bm f}_{b,m}^{[t]} \big) \notag \\
    &\quad +\frac{\chi_{k}}{2}\sum_{b,m}\big\| \tilde{\bm f}_{b,m}-\tilde{\bm f}_{b,m}^{[t]} \big\|_2^2 \notag \\
    & \triangleq \overline{I_k}^{[t]}(\{\tilde{\bm f}_{b,m}\}) . \label{eqq22}
\end{align}
\end{proposition}

\begin{IEEEproof}
    Please refer to Appendix~\ref{Appdix1}.
\end{IEEEproof}

In the remainder of this section, we focus on the regime $p\ge 2$. Besides facilitating the surrogate construction in Proposition~\ref{prop1}, this regime is also practically relevant. Specifically, $p\ge 2$ corresponds to moderately-to-highly directional elements, for which element-wise boresight steering can effectively exploit geometry selectivity and translate orientation control into tangible SINR/rate gains.  In contrast, when $p<2$ the power pattern becomes relatively broad and the performance is less sensitive to boresight alignment, so the benefit of element-wise orientation control is typically less pronounced. 
Using the bounds in Proposition~\ref{prop1} and relaxing the unit-norm constraints in \eqref{normCons} to their convex hull, we obtain the following joint convex surrogate problem at iteration $t$:
\begin{align}
    \label{prosurr_joint}
    \max_{\{\tilde{\bm{f}}_{b,m}\},\gamma,\{z_k\}} \, & \log_2(1+\gamma)  \\
    \text{s.t.} \qquad
    & \underline{S_k}^{[t]}(\{\tilde{\bm f}_{b,m}\}) \geq \frac{1}{2}\Big(\frac{z_k^{[t]}}{\gamma^{[t]}}\gamma^2+\frac{\gamma^{[t]}}{z_k^{[t]}}z_k^2\Big), \,\forall k, \\
    & \overline{I_k}^{[t]}(\{\tilde{\bm f}_{b,m}\}) + \sigma_k^2 \leq z_k,\,\, \forall k,\\
    & \cos(\theta_{\max}) \leq \tilde{\bm f}_{b,m}^T \bm{e}_z \leq 1,  \quad \forall b,m, \label{spc}\\
    & \|\tilde{\bm f}_{b,m}\|_2^2 \leq 1, \quad \forall b,m.
\end{align}
Problem~\eqref{prosurr_joint} can be efficiently solved by standard solvers such as CVX. Based on this surrogate, we develop an SCA-based algorithm for antenna orientation updates, as summarized in Algorithm~\ref{alg:orientation_joint}. Since the convex surrogate problem \eqref{prosurr_joint} relaxes the unit-norm constraints \eqref{normCons} by their convex hull $\|\tilde{\bm f}_{b,m}\|_2^2\le 1$, the orientation vectors obtained from Algorithm~\ref{alg:orientation_joint} may not have unit norm. We therefore apply a normalization step to enforce the unit-norm constraints before the subsequent beamformer update. 

\begin{algorithm}[t] \small
\caption{SCA-Based Algorithm for RA Optimization}
\label{alg:orientation_joint}
\begin{algorithmic}[1]
\REQUIRE Beamformers $\{\bm{w}_{k,b}\}$, initial orientations $\{\tilde{\bm{f}}_{b,m}^{[0]}\}$, tolerance $\epsilon_{\rm SCA}>0$, and maximum iterations $T_{\rm SCA}$
\ENSURE Optimized antenna orientations
\STATE Compute initial $\gamma^{[0]}$ and $\{z_k^{[0]}\}$;
\STATE Set $t \gets 0$;
\REPEAT
    \STATE Construct surrogate functions $\{\underline{S_k}^{[t]}\}$ and $\{\overline{I_k}^{[t]}\}$;
    \STATE Solve the convex surrogate problem \eqref{prosurr_joint} to obtain $\{\tilde{\bm f}_{b,m}^{[t+1]}\}$, $\gamma^{[t+1]}$, and $\{z_k^{[t+1]}\}$;
    \STATE Set $t \gets t + 1$;
\UNTIL{$t \ge T_{\rm SCA}$ or $\vert\gamma^{[t]} - \gamma^{[t-1]}\vert \le \epsilon_{\rm SCA}$}
\end{algorithmic}
\end{algorithm}

\begin{lemma}[Diagonal scaling induced by normalization]
\label{lem:diag_scaling}
Let $\{\tilde{\bm f}_{b,m}^{*}\}$ be any feasible solution to problem \eqref{prosurr_joint} and define the normalized boresights $\tilde{\bm f}_{b,m}^{[t+1]}\triangleq \tilde{\bm f}_{b,m}^{*}/\|\tilde{\bm f}_{b,m}^{*}\|_2, \, \forall b,m$. Under the channel model \eqref{eq:los_power}--\eqref{eq:nlos_coeff}, for every user $k$, the resulting stacked channel vector satisfies
\begin{align}
    \bm h_k\!\left(\{\tilde{\bm f}_{b,m}^{[t+1]}\}\right)
    = \bm D\,\bm h_k\!\left(\{\tilde{\bm f}_{b,m}^{*}\}\right),
\end{align}
where $\bm D\succeq \bm I$ is a diagonal matrix aligned with the stacked channel vector 
$\bm h_k=[\bm h_{k,1}^T,\ldots,\bm h_{k,B}^T]^T$. Specifically,
\begin{align}
    \bm D
    =\operatorname{diag}\big(
    \alpha_{1,1},\ldots,\alpha_{1,M},
    \ldots,
    \alpha_{B,1},\ldots,\alpha_{B,M}
    \big),
\end{align}
with $\alpha_{b,m}\triangleq \|\tilde{\bm f}_{b,m}^{*}\|_2^{-p}$.
\end{lemma}

\begin{IEEEproof}
Since $\theta_{\max}\in\big[0,\tfrac{\pi}{2}\big)$, we have $\cos(\theta_{\max})>0$, and thus constraint~\eqref{spc} ensures $\|\tilde{\bm f}_{b,m}^{*}\|_2 \neq 0$. Fix any unit direction vector $\bm s$ and define $\bm v\triangleq \bm R_b^T\bm s$. Then,
\begin{align}
    \big[(\bm R_b\tilde{\bm f}_{b,m}^{[t+1]})^T\bm s\big]_+^{p}
    &=\left[\frac{\tilde{\bm f}_{b,m}^{*T}\bm v}{\|\tilde{\bm f}_{b,m}^{*}\|_2}\right]_+^{p}
    =\|\tilde{\bm f}_{b,m}^{*}\|_2^{-p}\,[\tilde{\bm f}_{b,m}^{*T}\bm v]_+^{p} \notag\\
    &=\alpha_{b,m}\,\big[(\bm R_b\tilde{\bm f}_{b,m}^{*})^T\bm s\big]_+^{p},
    \label{eq:diag_scale_key2}
\end{align}
where $\alpha_{b,m}\triangleq \|\tilde{\bm f}_{b,m}^{*}\|_2^{-p}$. Moreover, by \eqref{eq:los_power} and \eqref{eq:nlos_coeff}, $\tilde{\bm f}_{b,m}$ affects the channel only through these directional-gain factors, whereas all distance-dependent terms and the corresponding phase rotations remain unchanged. Combining this observation with \eqref{eq:diag_scale_key2} yields
\begin{align}
    h_{k,b,m}\!\left(\{\tilde{\bm f}_{b,m}^{[t+1]}\}\right)
    =\alpha_{b,m}\,h_{k,b,m}\!\left(\{\tilde{\bm f}_{b,m}^{*}\}\right), \forall k,b,m.
\end{align}
Stacking over all elements gives $\bm h_k(\{\tilde{\bm f}_{b,m}^{[t+1]}\})=\bm D\,\bm h_k(\{\tilde{\bm f}_{b,m}^{*}\})$, where $\bm D$ is diagonal with diagonal entries $\{\alpha_{b,m}\}$. Finally, since problem \eqref{prosurr_joint} enforces $\|\tilde{\bm f}_{b,m}^{*}\|_2^2\le 1$, we have $\alpha_{b,m}\ge 1$ and hence $\bm D\succeq \bm I$.
\end{IEEEproof}

Lemma~\ref{lem:diag_scaling} reveals that the normalization step is benign, as it does not weaken the per-element channel magnitudes and only re-weights the stacked channel by a non-attenuating diagonal scaling. This property will be used later to show that normalization preserves the monotonicity of the AO updates, and hence does not affect the convergence.

\subsection{Overall Algorithm, Convergence, and Complexity}
Before running the AO iterations, we initialize the RA orientations using a best-of-$N_{\rm cand}$ strategy. Specifically, $N_{\rm cand}$ feasible orientation profiles are randomly generated within the spherical-cap constraint, and the one yielding the largest minimum user rate after SOCP-based beamforming is selected as $\{\tilde{\bm f}_{b,m}^{[0]}\}$. To sum up, Algorithm~\ref{AO-Alg} summarizes the proposed AO-based design for jointly optimizing the RA orientations and the downlink beamformers. At each outer iteration, we first update the RA orientations with the beamformers fixed by invoking Algorithm~\ref{alg:orientation_joint}, and then normalize the resulting orientation vectors to satisfy the unit-norm constraints. With these updated orientations fixed, we subsequently update the beamformers by solving the SOCP \eqref{SOCPBEAM}.

\begin{algorithm}[t] \small
    \caption{The Overall AO-Based Algorithm}
    \label{AO-Alg}
    \begin{algorithmic}[1]
        \REQUIRE Initial orientations $\{\tilde{\bm f}_{b,m}^{[0]}\}$, tolerance $\epsilon_{\rm AO}>0$, maximum number of iterations $T_{\rm AO}$
        \ENSURE Optimized beamformers and RA orientations
        \STATE With the initial RA orientations $\{\tilde{\bm f}_{b,m}^{[0]}\}$ fixed, obtain the initial beamformers $\{\bm w_{k,b}^{[0]}\}$ by solving Problem \eqref{SOCPBEAM};
        \STATE Compute the minimum user rate $R_{\min}^{[0]}$;
        \STATE Set $t \gets 0$;
        \REPEAT
            \STATE With $\{\bm w_{k,b}^{[t]}\}$ fixed, obtain a relaxed solution to RA orientations $\{\tilde{\bm f}_{b,m}^{*}\}$ using Algorithm~\ref{alg:orientation_joint};
            \STATE Normalize $\tilde{\bm f}_{b,m}^{[t+1]} = \tilde{\bm f}_{b,m}^{*}/\|\tilde{\bm f}_{b,m}^{*}\|_2$, $\forall b,m$;
            \STATE With $\{\tilde{\bm f}_{b,m}^{[t+1]}\}$ fixed, update the beamformers $\{\bm w_{k,b}^{[t+1]}\}$ by solving Problem \eqref{SOCPBEAM};
            \STATE Compute the minimum user rate $R_{\min}^{[t+1]}$;
            \STATE Set $t \gets t+1$;
        \UNTIL{$t \ge T_{\rm AO}$ or $\big|R_{\min}^{[t]} - R_{\min}^{[t-1]}\big| \le \epsilon_{\rm AO}$}
    \end{algorithmic}
\end{algorithm}

\begin{proposition}[Monotone convergence of Algorithm~\ref{AO-Alg}]
\label{prop:AO_convergence}
The minimum user rate $R_{\min}^{[t]}$ is monotonically non-decreasing with the iteration index $t$ of Algorithm~\ref{AO-Alg}, and hence is guaranteed to converge.
\end{proposition}

\begin{IEEEproof}
    Consider the $t$-th outer iteration of Algorithm~\ref{AO-Alg}. With fixed beamformers $\{\bm w_{k,b}^{[t]}\}$ and unit-norm orientations $\{\tilde{\bm f}_{b,m}^{[t]}\}$, the achieved worst-user SINR is $\gamma^{[t]}$. First, with $\{\bm w_{k,b}^{[t]}\}$ fixed, we update the RA orientations by solving the convex surrogate problem \eqref{prosurr_joint}. By construction, the surrogate inequalities in problem \eqref{prosurr_joint} are tight at the current point $\{\tilde{\bm f}_{b,m}^{[t]}\}$, and $(\{\tilde{\bm f}_{b,m}^{[t]}\},\gamma^{[t]},\{z_k^{[t]}\})$ satisfies all the surrogate constraints. Hence, $\{\tilde{\bm f}_{b,m}^{[t]}\}$ is a feasible point of problem \eqref{prosurr_joint}, and thus Algorithm~\ref{alg:orientation_joint} does not decrease the achievable worst-user SINR. Let $\gamma_{\text{sca}}$ and $\{\tilde{\bm f}_{b,m}^{*}\}$ denote, respectively, the achieved SINR value and the resulting orientation update returned by Algorithm~\ref{alg:orientation_joint}. It follows that
    \begin{align}
        \gamma_{\text{sca}} \ge \gamma^{[t]}.
    \end{align}
    We then normalize $\{\tilde{\bm f}_{b,m}^{*}\}$ to enforce the unit-norm constraints: $\tilde{\bm f}_{b,m}^{[t+1]}=\tilde{\bm f}_{b,m}^{*}/\|\tilde{\bm f}_{b,m}^{*}\|_2,\,\, \forall b,m.$ By Lemma~\ref{lem:diag_scaling}, we can construct a diagonal matrix $\bm D\succeq \bm I$ such that, for all $k$,
    \begin{align}
        \bm h_k\!\left(\{\tilde{\bm f}_{b,m}^{[t+1]}\}\right)=\bm D\,\bm h_k\!\left(\{\tilde{\bm f}_{b,m}^{*}\}\right).
    \end{align}
    Define the scaled beamformers $\bar{\bm w}_k\triangleq \bm D^{-1}\bm w_k^{[t]}$. Then, for any $k$ and $j$,
    \begin{align}
        \bm h_k\!\left(\{\tilde{\bm f}_{b,m}^{[t+1]}\}\right)^H \bar{\bm w}_j
        &=\left(\bm D\,\bm h_k\!\left(\{\tilde{\bm f}_{b,m}^{*}\}\right)\right)^H \left(\bm D^{-1}\bm w_j^{[t]}\right) \notag\\
        &=\bm h_k\!\left(\{\tilde{\bm f}_{b,m}^{*}\}\right)^H \bm w_j^{[t]}.
    \end{align}
    Hence, all desired-signal and interference terms are exactly preserved under $(\{\tilde{\bm f}_{b,m}^{[t+1]}\},\{\bar{\bm w}_k\})$, and thus the achieved minimum user SINR remains the same as that attained by $(\{\tilde{\bm f}_{b,m}^{*}\},\{\bm w_k^{[t]}\})$, i.e., $\gamma_{\text{sca}}$. Moreover, since $\bm D^{-1}\preceq \bm I$, the per-AP power constraints remain satisfied:
    \begin{align}
        \sum_{k=1}^{K}\|\bar{\bm w}_{k,b}\|_2^2 \le \sum_{k=1}^{K}\|\bm w_{k,b}^{[t]}\|_2^2 \le P_{\max},\quad \forall b.
    \end{align}
    Therefore, $(\{\tilde{\bm f}_{b,m}^{[t+1]}\},\{\bar{\bm w}_k\})$ is feasible for problem \eqref{SOCPBEAM} with the orientations fixed at $\{\tilde{\bm f}_{b,m}^{[t+1]}\}$, and achieves the SINR $\gamma_{\text{sca}}$. Solving \eqref{SOCPBEAM} optimally with $\{\tilde{\bm f}_{b,m}^{[t+1]}\}$ then yields an objective value no smaller than $\gamma_{\text{sca}}$, i.e.,
    \begin{align}
        \gamma^{[t+1]} \ge \gamma_{\text{sca}} \ge \gamma^{[t]}.
    \end{align}
    Consequently, $R_{\min}^{[t]}=\log_2(1+\gamma^{[t]})$ is non-decreasing with $t$. Moreover, $\{R_{\min}^{[t]}\}$ is upper-bounded due to the finite transmit power. Hence, by the monotone convergence theorem, $\{R_{\min}^{[t]}\}$ is guaranteed to converge.
\end{IEEEproof}

\textit{Complexity Analysis:} 
For the orientation optimization (Step~5 in Algorithm~\ref{AO-Alg}), the SCA procedure performs at most $T_{\mathrm{SCA}}$ iterations, and each iteration solves the convex surrogate problem \eqref{prosurr_joint} with worst-case complexity on the order of $\mathcal{O}\big((3MB+K)^{3.5}\big)$. The subsequent normalization is a simple per-element scaling and incurs only $\mathcal{O}(MB)$ operations, which is negligible compared with the cost of solving \eqref{prosurr_joint}. For the beamformer optimization (Step~7 in Algorithm~\ref{AO-Alg}), a bisection search with at most $T_{\mathrm{bis}}$ iterations is performed, and each iteration solves an SOCP whose worst-case complexity scales as $\mathcal{O}\big((BMK)^{3.5}\big)$. As a result, the overall complexity of Algorithm~\ref{AO-Alg} is $\mathcal{O}\big(T_{\rm AO}\big(T_{\mathrm{bis}} (BMK)^{3.5} + T_{\mathrm{SCA}} (3MB+K)^{3.5}\big)\big)$, where $T_{\rm AO}$ is the number of AO iterations.

\section{Proposed Low-Complexity Solution}

The AO algorithm in the previous section achieves high max--min rate performance, but it requires solving SOCP and SCA subproblems iteratively, which may result in high computational complexity. This motivates an efficient two-stage approach proposed in this section. Specifically, we first optimize the RA orientations to shape a fair radiation profile across users, and then, with the resulting geometry fixed, solve the max--min beamforming problem using the SOCP-based method in Section~\ref{sec:beamforming}.

We first construct a geometry-level performance metric that depends only on the RA orientations and the physical channel. For each user $k$, we define its aggregate channel gain as
\begin{align}
  \eta_k(\{\tilde{\bm f}_{b,m}\})
  \triangleq \bm{h}_k^H\bm{h}_k.
\end{align}
Intuitively, $\eta_k$ quantifies how much signal energy the RA-aided array can deliver to user $k$ under a given orientation geometry, irrespective of the specific beamforming vectors. To embed fairness directly at the geometry level, we adopt a logarithmic utility over the aggregate gains
\begin{align}
  U(\{\tilde{\bm f}_{b,m}\})
  \triangleq \sum_{k=1}^K \ln\big(\eta_k(\{\tilde{\bm f}_{b,m}\})+\varepsilon\big),
  \label{eq:log-geom-utility}
\end{align}
where $\varepsilon>0$ is a small regularization constant that prevents numerical issues when
$\eta_k$ is close to zero. This choice is in line with the well-known PF design principle: the marginal utility with respect to $\eta_k$ is
\begin{align}
  \frac{\partial U(\{\tilde{\bm f}_{b,m}\})}{\partial \eta_k}
  = \frac{1}{\eta_k(\{\tilde{\bm f}_{b,m}\})+\varepsilon},
\end{align}
which assigns larger weights to users with smaller aggregate gains and smaller weights to users with already strong gains. Consequently, maximizing $U(\{\tilde{\bm f}_{b,m}\})$ encourages the RA orientations to improve users with smaller aggregate channel gains, rather than only strengthening already favorable links. This is consistent with the max--min rate objective, since the worst-user rate is usually limited by users with weak effective channels. Therefore, although the utility in \eqref{eq:log-geom-utility} is not an exact equivalent of the max--min rate objective, it provides a low-complexity geometry-level proxy for constructing a more balanced channel realization before the subsequent SOCP-based max--min beamforming stage. Based on the utility in \eqref{eq:log-geom-utility}, the RA orientation design problem is formulated as
\begin{align}
  \max_{\{\tilde{\bm f}_{b,m}\}} \quad & U(\{\tilde{\bm f}_{b,m}\})
  \label{prob:PF-geom}\\
  \text{s.t.} \quad&
    \cos(\theta_{\max}) \leq \tilde{\bm f}_{b,m}^T \bm e_z \leq 1,  \quad \forall {b,m}, \label{cap1}\\
  & \|\tilde{\bm f}_{b,m}\|_2^2 = 1, \quad \forall b,m. \label{cap2}
\end{align}
Problem~\eqref{prob:PF-geom} remains nonconvex for two reasons: (i) the utility $U(\{\tilde{\bm f}_{b,m}\})$ couples all RA elements; and (ii) all RA orientation vectors $\{\tilde{\bm f}_{b,m}\}$ share a common nonconvex spherical-cap feasible set on the unit sphere, defined as 
\begin{align}
    \mathcal{C}_{\mathrm{cap}} \triangleq \big\{\bm x\in\mathbb{R}^3: \|\bm x\|_2=1,\ \bm x^T\bm e_z \ge \cos(\theta_{\max})\big\}. 
\end{align}
In what follows, we tackle Problem~\eqref{prob:PF-geom} using a Frank-Wolfe based algorithm. At iteration $t$, given a feasible orientation profile $\{\tilde{\bm f}^{[t]}_{b,m}\in\mathcal{C}_{\mathrm{cap}}\}$, we first compute the Euclidean gradient of the utility w.r.t. each RA orientation vector
\begin{align}
   \nabla_{\tilde{\bm{f}}_{b,m}} U = \sum_{k=1}^K \frac{2\Re\left((\bm{h}_{k}^{[t]})^H \nabla_{\tilde{\bm{f}}_{b,m}} \bm{h}_{k}^{[t]}\right)}{\eta_k^{[t]} + \varepsilon},
\end{align}
where the required channel derivatives $\{\nabla_{\tilde{\bm f}_{b,m}} \bm h_k^{[t]}\}$ are given in Appendix~\ref{Appdix1}. To respect the unit-norm constraint, we then project $\nabla_{\tilde{\bm{f}}_{b,m}} U$ onto the tangent space of the unit sphere at $\tilde{\bm f}^{[t]}_{b,m}$, obtaining the Riemannian gradient on the sphere
\begin{align}
    \label{RMG}
   \bm g^{[t]}_{b,m} &\triangleq \left(  \bm I - \tilde{\bm f}^{[t]}_{b,m}\big(\tilde{\bm f}^{[t]}_{b,m}\big)^{\!T} \right)(\nabla_{\tilde{\bm{f}}_{b,m}} U).
\end{align}
Next, for each element we select a ``target'' orientation on the spherical cap by maximizing the linearized utility along $\bm g^{[t]}_{b,m}$
\begin{align}
   \bm y^{[t]}_{b,m} 
   \in \arg\max_{\bm x\in\mathcal{C}_{\mathrm{cap}}} 
      \big\langle \bm g^{[t]}_{b,m}, \bm x \big\rangle,
\end{align}
whose solution can be obtained in closed-form as
\begin{align}
  \bm y^{[t]}_{b,m} 
  =
  \begin{cases}
    \widehat{\bm g}^{[t]}_{b,m}, & z^{[t]}_{b,m} \ge c_z, \\
    \sqrt{1-c_z^2}\bm v^{[t]}_{b,m}+ c_z \bm e_z, & z^{[t]}_{b,m} < c_z, \,\,\|\bm v^{[t]}_{b,m}\|_2\neq 0, \\
    \sqrt{1-c_z^2}\,\bm{u}+ c_z \bm e_z, & z^{[t]}_{b,m} < c_z, \,\,\|\bm v^{[t]}_{b,m}\|_2= 0,
  \end{cases}
  \label{eq:closed_form_solution_simplified}
\end{align}
where $\widehat{\bm g}^{[t]}_{b,m} \triangleq {\bm g}^{[t]}_{b,m}/\|{\bm g}^{[t]}_{b,m}\|_2$, $c_z \triangleq \cos(\theta_{\max})$, $z^{[t]}_{b,m} \triangleq \big(\widehat{\bm g}^{[t]}_{b,m}\big)^T \bm e_z$, $\bm v^{[t]}_{b,m} \triangleq (\widehat{\bm g}^{[t]}_{b,m} - z^{[t]}_{b,m}\bm e_z)/\vert| \widehat{\bm g}^{[t]}_{b,m} - z^{[t]}_{b,m}\bm e_z \vert|_2$, and $\bm u$ is any unit vector orthogonal to $\bm e_z$. The search direction for element $(b,m)$ is then defined as
\begin{align}
    \label{FWDir}
   \bm d^{[t]}_{b,m} \triangleq \bm y^{[t]}_{b,m} - \tilde{\bm f}^{[t]}_{b,m},
\end{align}
and all orientations are updated via a retraction
\begin{align}
   \tilde{\bm f}^{[t+1]}_{b,m}
   = \frac{\tilde{\bm f}^{[t]}_{b,m} + \rho^{[t]} \bm d^{[t]}_{b,m}}
           {\big\|\tilde{\bm f}^{[t]}_{b,m} + \rho^{[t]} \bm d^{[t]}_{b,m}\big\|_2},
   \label{eq:stage1_update}
\end{align}
where $\rho^{[t]}\in(0,1]$ is a stepsize that can be selected by a standard backtracking line search to guarantee a sufficient increase of the utility $U(\{\tilde{\bm f}_{b,m}\})$ at each accepted iteration. Given the geometry-optimized orientations from Stage~1, the effective channels $\{\bm h_k\}_{k=1}^K$ are fixed. In Stage~2, we solve the max--min fair beamforming problem \eqref{SOCPBEAM} once, using the SOCP-based bisection method described in Section~\ref{sec:beamforming}. Overall, the two-stage heuristic algorithm is summarized in Algorithm~\ref{AO-lowcom}.

\begin{algorithm}[t] \small
    \caption{Proposed Low-Complexity Algorithm}
    \label{AO-lowcom}
    \begin{algorithmic}[1]
    \REQUIRE Initial orientations $\{\tilde{\bm f}_{b,m}^{[0]}\}$, tolerance $\epsilon>0$, maximum number of iterations $T_{\rm FW}$.
    
    \nonumber \hspace{-\algorithmicindent}\textbf{Stage 1: Orientation Design}
    
    \STATE Set $t = 0$ and $U^{[0]} = U(\{\tilde{\bm f}_{b,m}^{[0]}\})$.
    \REPEAT
        \FOR{each antenna element $(b,m)$}
            \STATE Compute Riemannian gradient via \eqref{RMG}.
            \STATE Compute the target orientation via \eqref{eq:closed_form_solution_simplified}.
            \STATE Compute Frank-Wolfe search direction via \eqref{FWDir}.
        \ENDFOR
        \STATE Find the stepsize $\rho^{[t]}$ via Armijo backtracking.
        \FOR{each antenna element $(b,m)$}
            \STATE Update the antenna orientation $\{\tilde{\bm f}_{b,m}^{[t+1]}\}$ via \eqref{eq:stage1_update}.
        \ENDFOR
        \STATE Calculate $U^{[t+1]} = U(\{\tilde{\bm f}_{b,m}^{[t+1]}\})$.
        \STATE Set $t = t+1$.
    \UNTIL{$t \ge T_{\rm FW}$ or $\vert U^{[t]} - U^{[t-1]}\vert \le \epsilon$}
    
    \nonumber \hspace{-\algorithmicindent}\textbf{Stage 2: Beamforming Design}
    
    \STATE With antenna orientations fixed, obtain the optimal beamformers by solving problem \eqref{SOCPBEAM}.
    \end{algorithmic}
\end{algorithm}

For Algorithm~\ref{AO-lowcom}, the computational cost consists of two parts. In Stage~1, each FW iteration is dominated by the construction of the geometry-level objective and its gradients. Since the channel and gradient evaluations involve the LoS component and $Q$ scattered components for all $K$ users and $BM$ RA elements, the per-iteration complexity is on the order of $\mathcal{O}(KBM(Q+1))$. In Stage~2, with the RA orientations fixed, the SOCP-based max--min beamforming problem is solved once by bisection, with complexity $T_{\rm bis}\mathcal{O}((BMK)^{3.5})$. Therefore, the overall complexity of Algorithm~\ref{AO-lowcom} is approximated as
$\mathcal{O}\!\big(T_{\mathrm{FW}}\,KBM(Q+1) \;+\; T_{\mathrm{bis}}\,(BMK)^{3.5}\big)$, where $T_{\rm FW}$ and $T_{\rm bis}$ denote the numbers of FW iterations and bisection steps, respectively.

For clarity, Table~\ref{tab:complexity_comparison} summarizes the complexity comparison between the AO-based and low-complexity algorithms. Let $\mathcal{C}_{\rm BF}=T_{\rm bis}(BMK)^{3.5}$ denote the complexity order of one SOCP-based beamforming update with bisection. The computational saving mainly comes from three aspects. First, the low-complexity algorithm adopts a two-stage design and avoids the repeated orientation--beamforming alternation in the AO-based algorithm. Second, the SOCP-based beamforming problem is solved only once, rather than once in each AO iteration. Third, the FW orientation update admits a closed-form search direction on the spherical-cap constraint, thereby avoiding the repeated CVX-based SCA orientation subproblems.

\begin{table}[t] \footnotesize
\renewcommand{\arraystretch}{1.4}
\centering
\caption{Complexity Comparison}
\label{tab:complexity_comparison}
\begin{tabular}{lc}
\toprule[1.5pt]
\textbf{Algorithm} & \textbf{Computational Complexity} \\
\hline
AO-based algorithm 
& $\mathcal{O}\!\left(T_{\rm AO}\left(\mathcal{C}_{\rm BF}
+ T_{\rm SCA}(3BM+K)^{3.5}\right)\right)$ \\
Low-complexity algorithm 
& $\mathcal{O}\!\left(\mathcal{C}_{\rm BF}
+ T_{\rm FW}KBM(Q+1)\right)$ \\
\bottomrule[1.5pt]  
\end{tabular}
\end{table}

\begin{figure}[t]
	\begin{center}
		\includegraphics[width=0.48\textwidth]{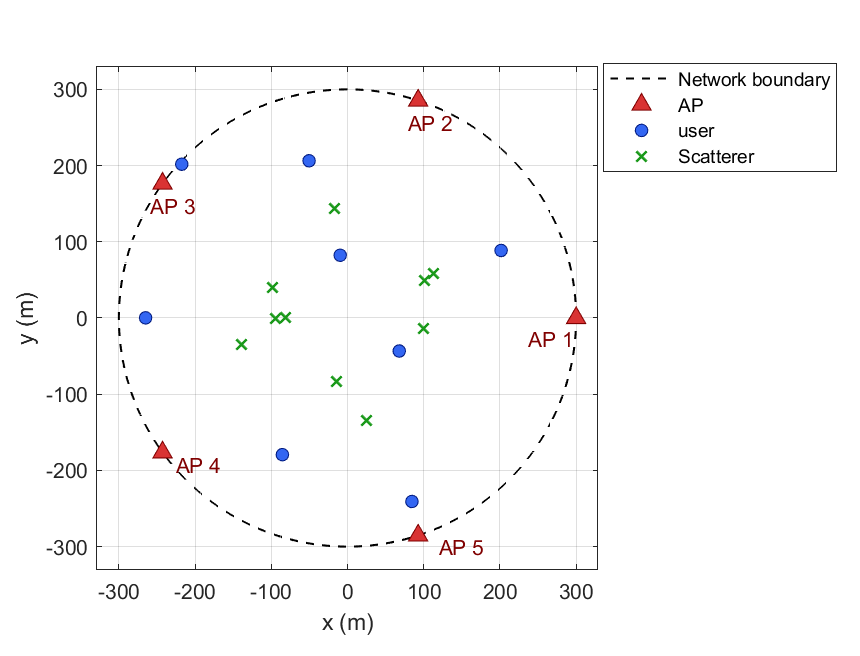}
		\caption{Network topology. Five APs are uniformly deployed on a circle with radius \(R_{\text{cov}} = 300\) m. The \(K = 8\) single-antenna users are uniformly distributed within the coverage disc.}
        \label{fig:toposetting}
	\end{center}
\end{figure}

\section{Simulation Results and Discussions}
Unless otherwise specified, we consider a cell-free network with \(B = 5\) APs uniformly deployed on a circle of radius \(R_{\text{cov}} = 300\) m at a height of \(h_{\text{AP}} = 30\) m, as illustrated in Fig. \ref{fig:toposetting}. Each AP is equipped with a UPA of \(M = 4\) antennas (\(M_x = 2\), \(M_y = 2\)), with an inter-element spacing of \(d = \lambda/2\). The UPA's boresight at each AP points toward the network center with a downtilt of \(5.7^\circ\). The system serves \(K = 8\) single-antenna users. All users are located at a height of \(h_{\text{U}} = 1.5\) m, and their horizontal positions are uniformly distributed within a disc of radius \(R_{\text{cov}}\). The carrier frequency is \(f_c = 2.4\) GHz, and the noise power at each user is set as \(\sigma^2 = -80\) dBm. The number of scatterers is set to \(Q=10\). The maximum numbers of AO and SCA iterations are set to $T_{\rm AO}=T_{\rm SCA}=200$, with stopping tolerances $\epsilon_{\rm AO}=\epsilon_{\rm SCA}=10^{-3}$. For the Frank--Wolfe update in the low-complexity algorithm, the Armijo parameter and backtracking factor are set to $c_{\rm A}=10^{-4}$ and $\tau_{\rm A}=0.5$, respectively.

\subsection{Convergence Behavior and Runtime Burden}

\begin{figure}[t]
	\begin{center}
		\includegraphics[width=0.48\textwidth]{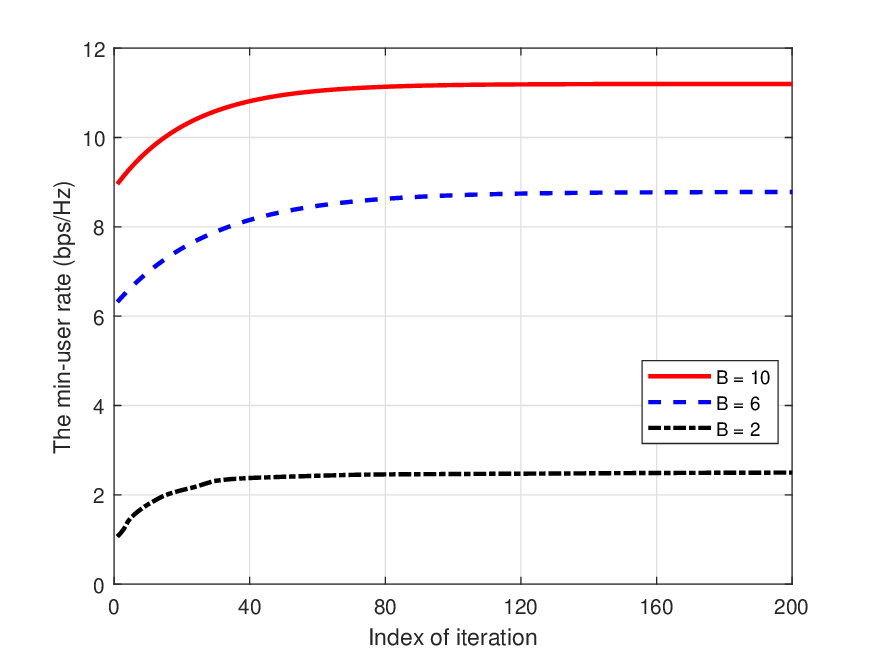}
		\caption{The convergence behavior of the proposed AO-based algorithm.}
        \label{fig:convergence}
	\end{center}
\end{figure}

Before the performance comparison, Fig.~\ref{fig:convergence} depicts the convergence behavior of the proposed AO-based algorithm under different AP deployments. Specifically, the transmit power budget per AP is set to $P_{\rm max}=15$~dBm and the antenna directivity factor is set to $p=5$. The figure plots the minimum user rate versus the AO iteration index for three representative numbers of APs, i.e., $B\in\{2,6,10\}$. In all cases, the minimum rate exhibits stable and non-decreasing behavior over iterations, which is consistent with the monotonicity property established in Proposition~2. This confirms that the proposed AO-based algorithm converges reliably for different AP deployment densities. Moreover, increasing $B$ improves the converged minimum rate, since more distributed APs provide stronger macro-diversity and more cooperative AP--user links. This allows the system to better support bottleneck users and improve the fairness-oriented minimum-rate performance.

To quantify the computational burden, the convergence tolerance of the AO-based algorithm is set to $\epsilon_{\rm AO}=10^{-3}$, and the simulations are conducted using MATLAB R2021b on a laptop equipped with a 2.4 GHz Intel(R) Core(TM) i7-13700H processor and 16 GB of RAM. For $B\in\{2,6,10\}$, the average runtimes of the AO-based algorithm are approximately $78$ s, $147$ s, and $376$ s, respectively. Under the same settings, the average runtimes of the proposed low-complexity algorithm are approximately $12$ s, $13$ s, and $15$ s, respectively. These results show that the low-complexity algorithm significantly reduces the runtime by avoiding repeated orientation--beamforming alternations and solving the SOCP-based beamforming problem only once. This runtime reduction becomes more pronounced as the number of APs increases.

\subsection{Performance Comparison}
For performance comparison, we consider the following benchmark schemes:
\begin{itemize}
    \item \textbf{Shared-orientation (AO):}  All RA elements within the same AP are constrained to share a common orientation, while different APs can have different common orientations. Under this AP-wise common-orientation constraint, the common orientations and transmit beamformers are optimized using the proposed AO-based framework.
    \item \textbf{Random antenna orientation:} The antenna orientations are randomly and uniformly drawn from their feasible sets to generate $30$ independent configurations. For each configuration, the max--min beamforming problem is solved using the SOCP-based method, and the one achieving the largest minimum user rate is reported. 
    \item \textbf{Isotropic antenna:} Each AP employs omnidirectional antennas (i.e., the antenna gain is orientation-independent, equivalently $p=0$ in Eq. \eqref{eq:RA_pattern}). The max--min beamforming problem is then solved using the SOCP-based method.
    \item \textbf{Fixed antenna orientation:} The antenna orientations are fixed to their initial values without optimization. Given this fixed geometry, the max--min beamforming problem is solved using the SOCP-based method.
\end{itemize}

\begin{figure}[t]
	\begin{center}
		\includegraphics[width=0.48\textwidth]{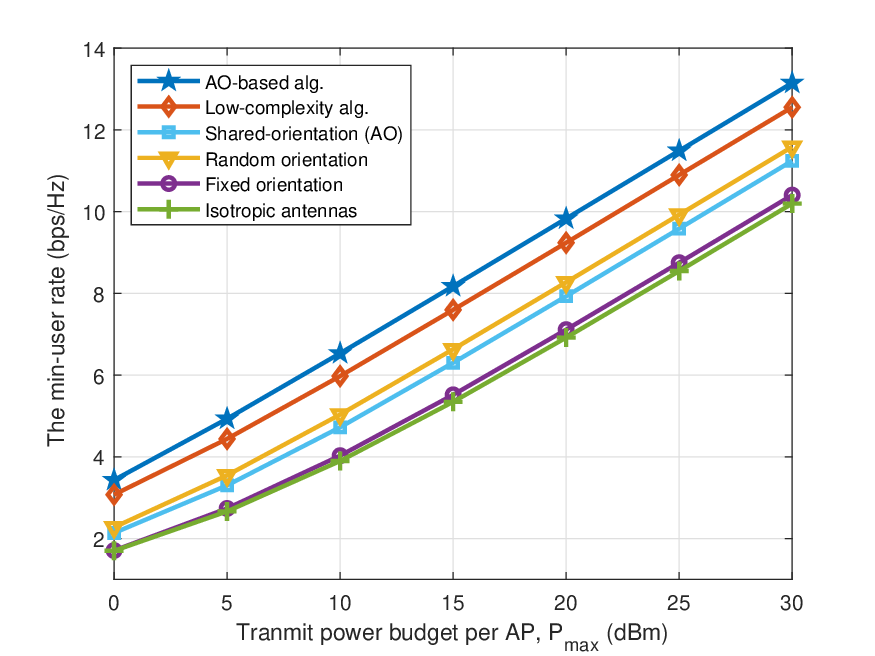}
		\caption{The min-user rate vs. the transmit power budget per AP.}
        \label{fig:ptrange}
	\end{center}
\end{figure}

Fig.~\ref{fig:ptrange} shows the average minimum user rate versus the transmit power budget per AP, where we set $p=5$ and $\theta_{\text{max}}=\pi/3$. The proposed AO-based and low-complexity algorithms consistently outperform all benchmarks over the entire power range, and their curves are nearly parallel to the baselines. For instance, at $P_{\max}=15$ dBm, the AO-based method improves the worst-user rate by approximately $24\%$, $49\%$, and $53\%$ over the random orientation, isotropic, and fixed-orientation baselines, respectively. The low-complexity algorithm also delivers notable gains of about $15\%$, $38\%$, and $43\%$ over the same benchmarks. The random-orientation baseline is also relatively competitive due to the best-of-multiple-trials selection, while the shared-orientation scheme provides an intermediate performance level because the AP-wise common-orientation constraint reduces the orientation flexibility while still allowing each AP to adapt its common radiation direction. These gains are attributed to RA orientation optimization, which leverages antenna directivity to reinforce weak AP--user links while suppressing dominant interference, thereby enabling more effective multi-AP cooperation and lifting the worst-user rate floor.

\begin{figure}[t]
	\centering
	\includegraphics[width=0.48\textwidth]{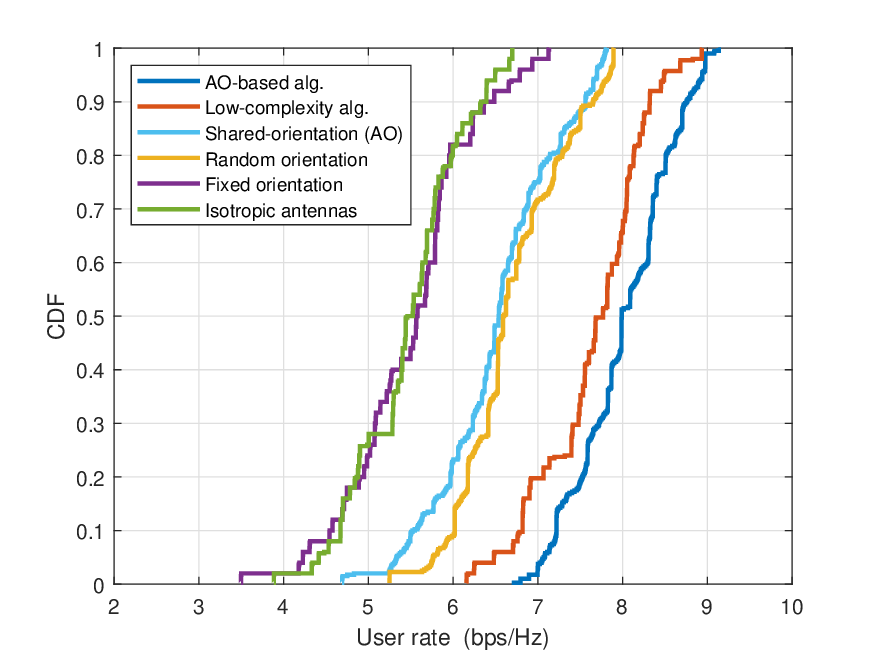}
    \caption{CDF of individual user rates under different schemes.}
	\label{rateCDF}
\end{figure}

To further characterize the rate distribution across users, Fig.~\ref{rateCDF} plots the cumulative distribution function (CDF) of individual user rates under $P_{\max}=15$ dBm. Compared with the benchmark schemes, the proposed AO-based algorithm shifts the CDF curve to the right, especially in the low-rate region. Most user rates achieved by the proposed AO-based and low-complexity algorithms are concentrated in a higher-rate range, whereas the fixed-orientation and isotropic-antenna baselines exhibit a larger portion of low-rate users. This indicates that the proposed RA orientation design improves not only the bottleneck-user performance but also the overall user-rate distribution. The proposed low-complexity algorithm closely follows the AO-based algorithm, showing that it preserves most of the performance gain. In addition, the random-orientation baseline remains competitive due to the best-of-multiple-trials selection, while the shared-orientation scheme still outperforms the fixed-orientation and isotropic-antenna baselines due to its optimized AP-wise common orientations.

\begin{figure}[t]
	\begin{center}
		\includegraphics[width=0.48\textwidth]{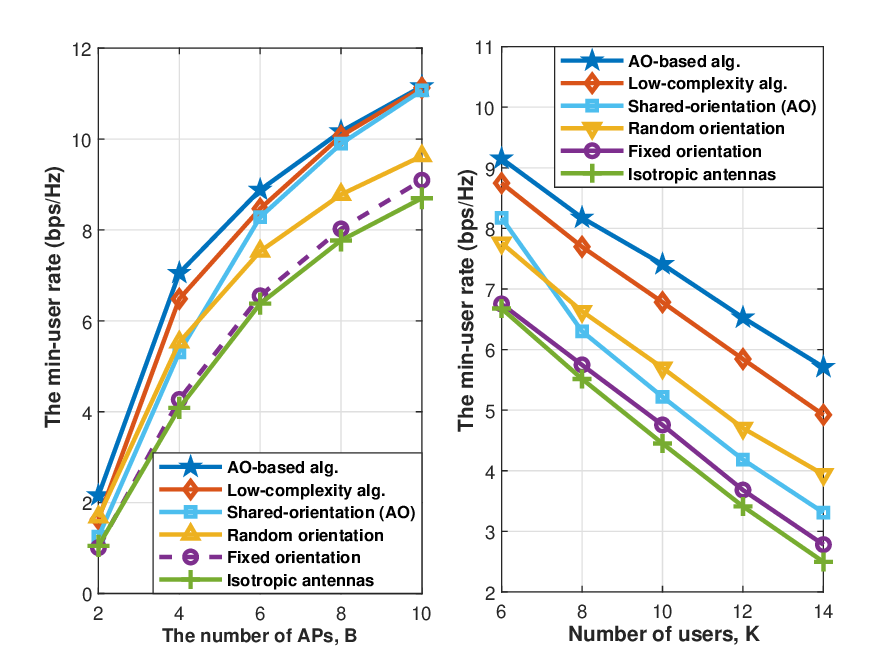}
		\caption{The min-user rate vs. (left) the number of APs and (right) the number of users.}
        \label{fig:density}
	\end{center}
\end{figure}

The left panel of Fig.~\ref{fig:density} shows the performance versus the number of APs, where we set $K=8$, $P_{\text{max}}=15$ dBm, $p=5$, and $\theta_{\text{max}}=\pi/3$. As the number of APs increases, all schemes achieve higher minimum rates due to the enhanced macro-diversity and the increased number of cooperative transmit antennas. More importantly, the performance gap between the RA-optimized schemes and the fixed-orientation benchmarks highlights the interplay between macro-diversity and antenna directivity: increasing the number of APs provides more distributed AP--user links, while RA orientation optimization determines how effectively these links are converted into fairness gains. The proposed AO-based algorithm consistently achieves the best performance over the entire AP range, while the proposed low-complexity algorithm closely approaches it, especially when $B$ becomes large. This shows that the geometry-guided low-complexity design can capture a large portion of the RA gain. In addition, the shared-orientation scheme also outperforms the fixed-orientation and isotropic-antenna baselines, demonstrating the effectiveness of the proposed AO framework even under the more practical AP-wise common-orientation constraint. Nevertheless, the fully element-wise RA design still provides the best performance due to its finer orientation flexibility. The right panel of Fig.~\ref{fig:density} plots the average minimum user rate versus the number of users with $B=5$. As the number of users increases, the minimum user rate decreases for all schemes, since more users compete for the available spatial resources and the max--min fairness requirement becomes more stringent. The proposed AO-based algorithm remains the best-performing scheme, and the proposed low-complexity algorithm follows it, showing the effectiveness of the proposed RA orientation design under heavier user loads. The shared-orientation scheme provides a meaningful gain over the fixed-orientation and isotropic-antenna baselines, but its performance decreases more rapidly as $K$ grows, because a single common orientation per AP becomes less flexible when serving more users with diverse channel directions.

\begin{figure}[t]
	\begin{center}
		\includegraphics[width=0.48\textwidth]{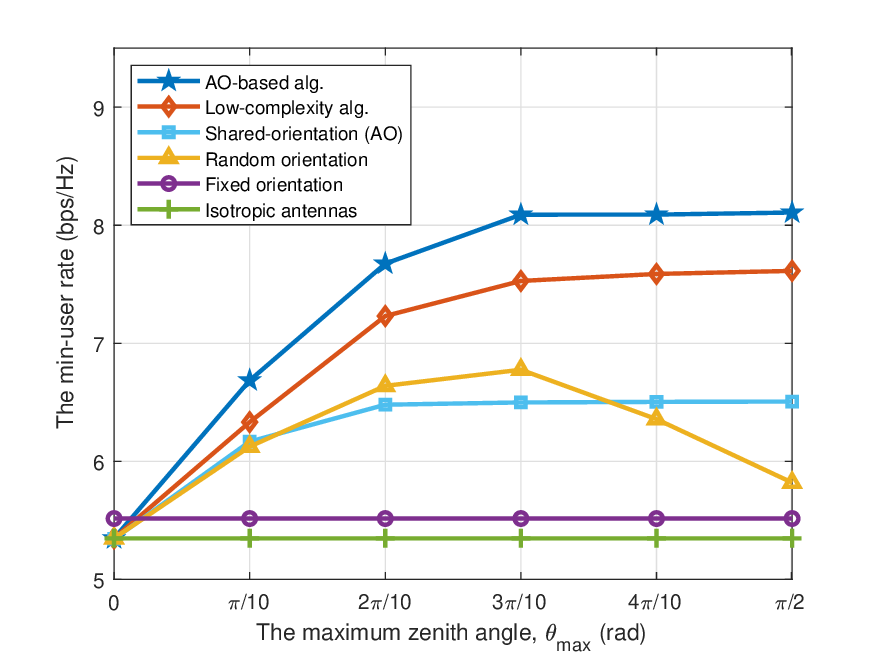}
		\caption{The min-user rate vs. the maximum zenith angle.}
        \label{fig:angleRange}
	\end{center}
\end{figure}

Fig.~\ref{fig:angleRange} shows the average minimum user rate versus the maximum zenith angle $\theta_{\max}$, where we set $P_{\text{max}} = 15$ dBm and $p=5$. As $\theta_{\max}$ increases, the proposed AO-based algorithm achieves a rapid performance improvement and then gradually saturates, suggesting that a moderate steering range already captures most of the RA orientation gain. The proposed low-complexity algorithm follows a similar trend but converges to a lower plateau, since it does not iteratively optimize beamforming and orientations as in the AO-based design. The shared-orientation scheme also benefits from the enlarged orientation range, but its gain saturates earlier due to the AP-wise common-orientation constraint. This indicates that shared orientation can capture part of the RA gain with a more practical control structure, while the fully element-wise RA design still provides the best performance due to its finer flexibility. In contrast, the random-orientation benchmark achieves limited gains and can even deteriorate for large $\theta_{\max}$, showing that a larger feasible orientation range is not inherently beneficial without principled optimization. The fixed-orientation and isotropic-antenna baselines remain unchanged, further confirming that the additional orientation flexibility is beneficial only when RA orientations are properly optimized.

\begin{figure}[t]
	\begin{center}
		\includegraphics[width=0.48\textwidth]{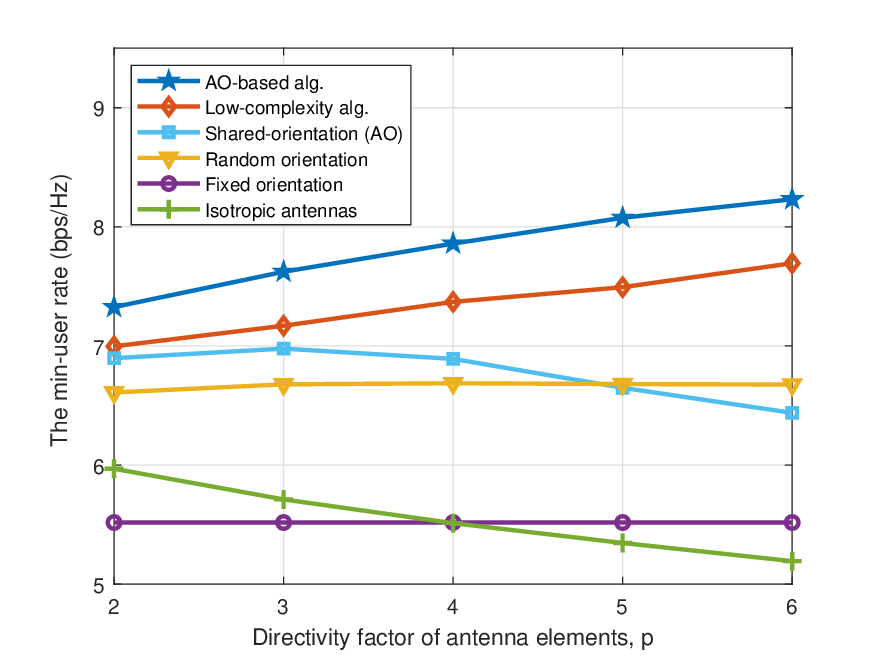}
		\caption{The min-user rate vs. the antenna directivity factor.}
        \label{fig:pdirRange}
	\end{center}
\end{figure}

\begin{figure}[t]
	\centering
	\includegraphics[width=0.48\textwidth]{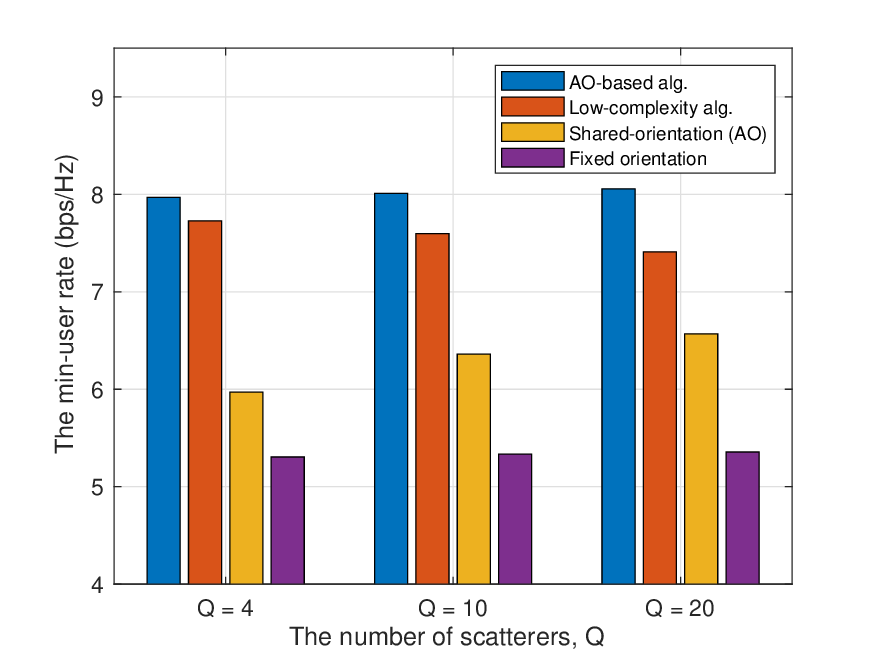}
    \caption{Performance comparison under different numbers of scatterers.}
	\label{fig:QRange}
\end{figure}

Fig.~\ref{fig:pdirRange} shows the average minimum user rate versus the antenna directivity factor $p$, where we set $P_{\text{max}}=15$ dBm, and $\theta_{\text{max}} = \pi/3$. The two proposed schemes benefit monotonically from increasing $p$, indicating that stronger antenna directivity can be effectively converted into fairness gains when the RA orientations are properly adjusted and jointly designed with beamforming. This observation further highlights the interplay between macro-diversity and antenna directivity, since stronger directivity can be fully exploited only when the effective distributed AP--user link qualities are properly shaped through RA orientation optimization. In contrast, the ``Fixed antenna orientation'' baseline degrades noticeably as $p$ increases, since a narrower main-lobe with misaligned boresights reduces the effective coverage of the worst users and exacerbates unfairness. The ``Isotropic antenna'' baseline is insensitive to $p$, as expected, while the ``Random antenna orientation'' scheme exhibits limited improvement and can even deteriorate for large $p$, suggesting that higher directivity is not inherently beneficial without principled orientation control. The shared-orientation scheme captures part of the RA gain under moderate directivity, but its performance degrades when $p$ becomes large because the narrower antenna main lobes make the AP-wise common-orientation constraint less flexible. These results also show that the observed RA gain is not tied to a single directivity setting within the adopted cosine-power model. Overall, the figure highlights that directivity is a double-edged sword: it amplifies the benefits of optimized RA designs, yet can be detrimental when orientations are not properly optimized. 

Next, to examine the impact of scattering conditions, Fig.~\ref{fig:QRange} compares the min-user rate under different numbers of scatterers, i.e., $Q\in\{4,10,20\}$, while keeping the other simulation parameters unchanged. The performance variation with $Q$ is moderate, suggesting that the main performance trend is not overly sensitive to the considered scattering conditions. More importantly, the proposed AO-based and low-complexity algorithms consistently outperform the fixed-orientation baseline for all considered values of $Q$, confirming that RA orientation optimization remains effective under different scattering environments. The shared-orientation scheme also provides a clear gain over the fixed-orientation baseline, indicating that AP-wise common orientation control can exploit part of the propagation-direction information. Nevertheless, the fully element-wise RA design achieves the best performance due to its finer orientation flexibility.

\begin{figure}[t]
	\centering
	\includegraphics[width=0.48\textwidth]{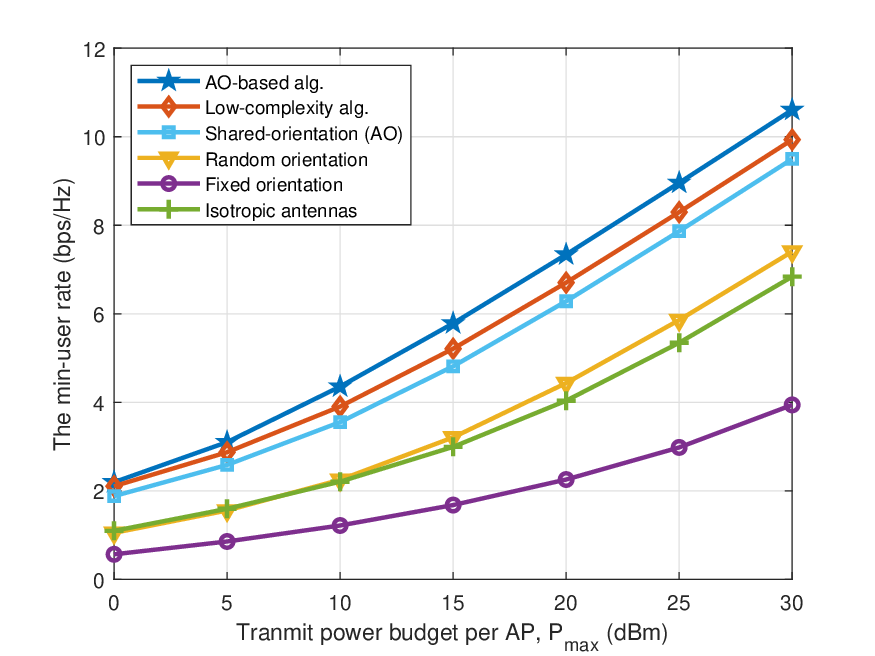}
    \caption{Min-user rate versus the transmit power budget per AP under the clustered edge-street user distribution.}
	\label{fig:clustered_ptrange}
\end{figure}

Finally, to examine the impact of user geometry, we consider a clustered edge-street scenario, which jointly captures clustered-user and cell-edge-user conditions. In this setup, users are concentrated in a street-like region near the service boundary, and the street is placed between two neighboring APs. Compared with the default uniform distribution, this scenario leads to weaker and more imbalanced AP--user links and is therefore more challenging for max--min fair transmission. Fig.~\ref{fig:clustered_ptrange} shows the corresponding min-user rate versus the transmit power budget per AP. Compared with the default uniform-user case in Fig.~\ref{fig:ptrange}, all schemes experience noticeable performance degradation under this clustered edge-street scenario. Nevertheless, the proposed AO-based algorithm still achieves the best performance, and the proposed low-complexity algorithm follows it over the whole power range, showing that RA orientation optimization remains effective under clustered cell-edge conditions. In addition, the shared-orientation scheme becomes relatively competitive because an AP-wise common orientation can partially match the dominant user direction, although the fully element-wise RA design still achieves the highest performance due to its finer orientation flexibility. These results further support the main insight of this paper that macro-diversity and antenna directivity play complementary roles in cell-free MIMO systems. Distributed APs provide multiple cooperative links, while RA orientation optimization reshapes the effective AP--user link qualities, enabling the network to better support bottleneck users under both dispersed and clustered user distributions.

\section{Conclusion}

This paper studied max--min fair downlink transmission in cell-free networks enhanced by RAs. We formulated a max--min rate maximization problem that jointly optimizes the network-wide beamformers and the element-wise RA orientations under per-AP power budgets and spherical-cap steering constraints. To tackle the resulting nonconvex design, we developed an AO framework, where the beamforming subproblem is solved optimally via bisection with SOCP feasibility checks, and the orientation subproblem is handled by an SCA procedure with a relaxation of the unit-norm constraints followed by a normalization step that preserves feasibility and the achieved minimum-rate value. In addition, to reduce computational complexity, we proposed an efficient two-stage scheme that first designs the RA orientations by maximizing a proportional-fair log-utility using manifold-aware Frank--Wolfe updates, and then computes the beamformers once using the SOCP-based max--min design.

Simulation results demonstrated that incorporating RA orientation optimization consistently improves the worst-user rate over beamforming-only baselines and random orientation schemes. A moderate steering range was shown to capture most of the achievable gain, supporting practical actuation limits. The results also highlighted that antenna directivity is beneficial for fairness only when the boresights are properly optimized, whereas misaligned highly directional elements can degrade the worst-user performance. Overall, the proposed RA-aware designs provide an effective and implementation-friendly means to strengthen disadvantaged links and enhance service uniformity in cell-free deployments. An important direction for future work is to consider more practical orientation-control architectures, such as group-wise or subarray-wise rotation control. Another is to develop more realistic hardware-aware and robust RA designs that account for practical impairments, channel uncertainty, and other higher-order effects. In addition, it is also important to develop two-timescale RA and beamforming designs to account for the practical response-speed limitation of antenna rotation, where RA orientations are updated on a slow time scale according to slowly varying channel/geometry information, while digital beamformers are adapted on a fast time scale according to short-term CSI.

\appendix

\subsection{Proof of Proposition~\ref{prop1}}

\label{Appdix1}

\begin{figure*}[b] 
    \setlength{\arraycolsep}{5pt}
    \hrulefill
    \vspace*{4pt}
    \setlength{\arraycolsep}{0.0em}
    \begin{align}
    \nabla_{\tilde{\bm f}_{b,m}} h_{k,b,m} 
    &=  \sqrt{\beta_0\kappa_{\max}}\;
        \frac{p}{r_{k,b,m}}\,
        \big[(\bm R_b\tilde{\bm f}_{b,m})^{T}\bm s_{k,b,m}\big]_+^{\,p-1}\,
        \bm R_b^{T}\bm s_{k,b,m}\,
        e^{-j\frac{2\pi}{\lambda} r_{k,b,m}} \notag\\
    & + \sum_{q=1}^Q 
        \sqrt{\beta_0\kappa_{\max}}\sqrt{\tfrac{\zeta_q}{4\pi}}\;
        \frac{p}{\tilde r_{q,b,m}\,\hat r_{k,q}}\,
        \big[(\bm R_b\tilde{\bm f}_{b,m})^{T}\bm s_{q,b,m}\big]_+^{\,p-1}\,
        \bm R_b^{T}\bm s_{q,b,m}\,
        e^{-j\frac{2\pi}{\lambda}(\tilde r_{q,b,m}+\hat r_{k,q})+j\chi_q},
       \label{big1}\\[2mm]
    \nabla_{\tilde{\bm f}_{b,m}}^{2} h_{k,b,m}
    &= \sqrt{\beta_0\kappa_{\max}}\;
       \frac{p(p-1)}{r_{k,b,m}}\,
       \big[(\bm R_b\tilde{\bm f}_{b,m})^{T}\bm s_{k,b,m}\big]_+^{\,p-2}\,
       (\bm R_b^{T}\bm s_{k,b,m})(\bm R_b^{T}\bm s_{k,b,m})^{T}\,
       e^{-j\frac{2\pi}{\lambda} r_{k,b,m}} \notag\\
    & + \sum_{q=1}^{Q}
       \sqrt{\beta_0\kappa_{\max}}\sqrt{\tfrac{\zeta_q}{4\pi}}\;
       \frac{p(p-1)}{\tilde r_{q,b,m}\,\hat r_{k,q}}\,
       \big[(\bm R_b\tilde{\bm f}_{b,m})^{T}\bm s_{q,b,m}\big]_+^{\,p-2}\,
       (\bm R_b^{T}\bm s_{q,b,m})(\bm R_b^{T}\bm s_{q,b,m})^{T}\,
       e^{-j\frac{2\pi}{\lambda}(\tilde r_{q,b,m}+\hat r_{k,q})+j\chi_q}.
       \label{big2}
    \end{align}
\end{figure*}

By using the chain rule, the first-order gradients of the signal and interference powers in \eqref{SigPNoisP} w.r.t. each element-wise boresight vector $\tilde{\bm f}_{b,m}\in\mathbb{R}^3$ are given by
\begin{align}
    \nabla_{\tilde{\bm f}_{b,m}} S_k
    &=2\,\Re\!\left\{ w_{k,b,m}^{*}\,(\bm h_k^{H}\bm w_k)\,
    \nabla_{\tilde{\bm f}_{b,m}} h_{k,b,m}\right\}, \label{eq:grad_S_rank1}\\
    \nabla_{\tilde{\bm f}_{b,m}} I_k
    &=2\,\Re\!\Big\{\Big(\sum_{j\neq k} w_{j,b,m}^{*}\,(\bm h_k^{H}\bm w_j)\Big)
    \nabla_{\tilde{\bm f}_{b,m}} h_{k,b,m}\Big\}, \label{eq:grad_I_rank1}
\end{align}
where $\nabla_{\tilde{\bm f}_{b,m}} h_{k,b,m}$ is given in \eqref{big1} at the bottom of the next page. The corresponding second-order gradients are given by
\begin{align}
    \nabla_{\tilde{\bm f}_{b,m}}^{2} S_k
    &=
    2\,\Re\!\Big\{
    |w_{k,b,m}|^{2}\,
    \nabla_{\tilde{\bm f}_{b,m}} h_{k,b,m}
    \big(\nabla_{\tilde{\bm f}_{b,m}} h_{k,b,m}\big)^{H}
    \Big\} \notag\\
    &\,\,+
    2\,\Re\!\Big\{
    w_{k,b,m}^{*}\,(\bm h_k^{H}\bm w_k)\,
    \nabla_{\tilde{\bm f}_{b,m}}^{2} h_{k,b,m}
    \Big\}.
    \label{eq:hess_S_correct} \\
    \nabla_{\tilde{\bm f}_{b,m}}^{2} I_k
    &= \notag
    2\,\Re\!\big\{
    \sum_{j\neq k}|w_{j,b,m}|^{2}
    \nabla_{\tilde{\bm f}_{b,m}} h_{k,b,m}
    \big(\nabla_{\tilde{\bm f}_{b,m}} h_{k,b,m}\big)^{H}
    \big\} \notag\\
    &\,\,+
    2\,\Re\!\big\{
    \big(\sum_{j\neq k} w_{j,b,m}^{*}\,(\bm h_k^{H}\bm w_j)\big)
    \nabla_{\tilde{\bm f}_{b,m}}^{2} h_{k,b,m}
    \big\}.
    \label{eq:hess_I_correct}
\end{align}
where $\nabla_{\tilde{\bm f}_{b,m}}^{2} h_{k,b,m}$ is given in \eqref{big2} at the bottom of the next page. Due to the positive-part operator $[\cdot]_+$, when $p<2$, the functions $S_k$ and $I_k$ are not second-order Lipschitz smooth in $\tilde{\bm f}_{b,m}$, so one cannot guarantee a global quadratic lower bound with a finite curvature constant. For $p\ge 2$, the quadratic lower bound in Proposition~\ref{prop1} can be ensured by constructing a curvature constant $\xi_k$ such that $\|\nabla_{\tilde{\bm f}_{b,m}}^{2} S_k\|_2 \le \xi_k$ for all feasible $\tilde{\bm f}_{b,m}$ (equivalently, $\nabla_{\tilde{\bm f}_{b,m}}^{2} S_k \preceq \xi_k \bm I$). Since $\|\nabla_{\tilde{\bm f}_{b,m}}^{2} S_k\|_2 \le \|\nabla_{\tilde{\bm f}_{b,m}}^{2} S_k\|_F$, it suffices to upper-bound the Frobenius norm. Under the unit-norm constraint on $\tilde{\bm f}_{b,m}$, $[(\bm R_b\tilde{\bm f}_{b,m})^T\bm s]_+\le 1$, and the orthogonality of $\bm R_b$ ensures that $[\,\cdot\,]_+^{p-1}$ and $[\,\cdot\,]_+^{p-2}$ are globally bounded for $p\ge2$. Consequently, \(\|\nabla_{\tilde{\bm f}_{b,m}} h_{k,b,m}\|_2\) and \(\|\nabla_{\tilde{\bm f}_{b,m}}^{2} h_{k,b,m}\|_F\) are bounded above. Substituting these bounds into \eqref{eq:hess_S_correct} yields an explicit upper bound for \(\|\nabla_{\tilde{\bm f}_{b,m}}^{2} S_k\|_F\), from which a finite constant \(\xi_k\) can be constructed. Applying the same reasoning to \eqref{eq:hess_I_correct} analogously provides a finite \(\chi_k\) for \(I_k\).

\end{document}